\documentclass[hidelinks,onefignum,onetabnum]{siamart251216}

\usepackage{lipsum}
\usepackage{amsfonts}
\usepackage{graphicx}
\usepackage{epstopdf}
\usepackage{algorithmic}
\usepackage{tikz}
\usetikzlibrary{arrows.meta}
\ifpdf
  \DeclareGraphicsExtensions{.eps,.pdf,.png,.jpg}
\else
  \DeclareGraphicsExtensions{.eps}
\fi
\usepackage[caption=false]{subfig}
\usepackage{mathtools}
\usepackage{enumitem}
\usepackage{booktabs}
\usepackage{array}
\usepackage{multirow}

\graphicspath{{graphics/}}

\newsiamremark{remark}{Remark}
\newsiamremark{hypothesis}{Hypothesis}
\crefname{hypothesis}{Hypothesis}{Hypotheses}
\newsiamthm{claim}{Claim}
\newsiamremark{fact}{Fact}
\crefname{fact}{Fact}{Facts}

\headers{Sharp--diffuse interface model}{R. Ramani}

\title{A sharp--diffuse interface model for intermittent and isolated topological transitions}

\author{Raaghav Ramani\thanks{Center for Nonlinear Studies, Los Alamos National Laboratory, Los Alamos, NM 87545 
  (\email{raaghav.ramani@outlook.com}).}}

\usepackage{amsopn}

\newsiamthm{assumption}{Assumption}
\crefname{assumption}{assumption}{assumptions}
\Crefname{assumption}{Assumption}{Assumptions}

\newsiamthm{example}{Example}
\crefname{example}{example}{examples}
\Crefname{example}{Example}{Examples}

\def\tmax{{t}_{\mathsf{max}}}

\def\dd#1{{\,\mathrm{d}#1}}

\newcommand{\gradd}{\operatorname{grad}}
\newcommand{\trace}{\operatorname{tr}}
\newcommand{\dist}{\operatorname{dist}}
\newcommand{\diam}{\operatorname{diam}}
\def\p{\partial}
\def\eps{\varepsilon}
\def\R{\mathbb{R}}
\def\S{\mathbb{S}}
\def\O{\mathcal{O}}
\def\Ngrid{{N}_{\mathsf{grid}}}

\def\Eng{{E}}
\def\Feng{\mathcal{F}}
\def\Mass{\mathcal{M}}
\def\sigmah{\widehat{\sigma}}
\def\etah{{\chi}}

\def\Gammabf{\mathbf{\Gamma}}
\def\reach{\operatorname{reach}}
\def\Clyr{C_{\mathsf{layer}}}
\def\Tlyr{\mathcal{T}}

\def\OmegaCH{{\Omega}}
\def\OmegaCHm{{\Omega}_{k}^{\eps}}
\def\Z{\mathcal{Z}}
\def\Zdfct{\mathcal{Z}_{k}^{\eps}}
\def\cbuf{c_{\mathsf{buf}}}
\def\ain{\mathfrak{a}}
\def\smax{s_{\mathsf{max}}}

\def\Gker{\mathcal{G}}

\def\Sop{\mathcal{S}}

\ifpdf
\hypersetup{
  pdftitle={A sharp--diffuse interface model for intermittent and isolated topological transitions},
  pdfauthor={R. Ramani}
}
\fi

\begin{document}

\maketitle

\begin{abstract}
We propose a hybrid sharp--diffuse interface representation for
modeling intermittent and isolated topological transitions during 
Cahn--Hilliard phase coarsening.
Away from topological events, the evolution is approximated by the
Mullins--Sekerka sharp-interface limit system and computed using a boundary
integral formulation.  When diffuse transition layers overlap, a novel 
\emph{interface surgery} algorithm resolves topology changes through a
localized Cahn--Hilliard pseudo-time evolution, after which the sharp-interface 
calculation is resumed.  We develop
the mathematical formulation underlying this decomposition, present
a simple two-dimensional numerical implementation, and simulate a
mass-exchange problem with interface coalescence. 
By localizing the diffuse evolution to topological events, the method
achieves a speedup of two to three orders of magnitude over
conventional diffuse-interface simulations.
\end{abstract}




\section{Introduction}

The description of an interface as a thin transition layer separating
two distinct phases traces back to the foundational work of van der
Waals and Korteweg. Building on this diffuse-interface picture, Cahn
and Hilliard~\cite{CaHi1958} developed a variational model for phase
separation and coarsening in binary alloys subjected to rapid cooling,
or quenching. The quench renders an initially homogeneous mixture
thermodynamically unstable: it quickly \emph{separates} into nearly pure
phases divided by thin transition layers, and then \emph{coarsens} more slowly
as diffusion increases the characteristic size of these regions.
The \emph{Cahn--Hilliard equation} (see \eqref{CH} below), which is
the fundamental diffuse-interface model studied in this work, represents this
two-stage evolution as a mass-conserving gradient flow of the
Ginzburg--Landau free energy. This energetic framework is highly
flexible and has been applied across a wide range of disciplines
\cite{AnMcWh1998,Chen2002,Kim2012,PaPa2025}.

The Cahn--Hilliard equation and related systems 
have also been studied extensively in the mathematical literature. 
For a comprehensive treatment, we refer to the monograph of
Miranville~\cite{Miranville2019}.
Elliott and Zheng~\cite{ElZh1986} established existence, uniqueness, and
regularity for the constant-mobility case with a
polynomial free-energy density, while 
Nicolaenko--Scheurer--Temam~\cite{NiScTe1989} analyzed the long-time behavior. 
In the phase-coarsening regime of interest here, heuristic scaling
predicts that the coarsening length scale $\ell(t)$, measuring
the typical size of the phase domains, grows according to the slow law
$\ell(t)\sim t^{1/3}$ \cite{KoOt2002}.

During phase coarsening, interfaces can undergo topology changes, which 
we assume are intermittent and isolated in a sense that will be made 
precise below. These topological events are governed by diffuse mixing on
spatial and temporal scales of order $\O(\eps)$ and $\O(\eps^3)$,
respectively, where $\eps$ denotes the transition-layer thickness. 
In practical applications, $\eps$ is much
smaller than the macroscopic length scale $L$ of the interfacial
evolution. Direct diffuse-interface simulations must resolve both the transition
layers and the topological events with sufficient spatiotemporal
resolution. High-order spatial discretizations
\cite{LiSh2003}, energy-stable semi-implicit time integrators
\cite{ShXuYa2018,ShXuYa2019}, and sophisticated adaptive
mesh-refinement strategies \cite{WiKiLo2007} have enabled simulations of
complex multiscale phenomena, but entail significant computational cost
and implementation complexity.

These computational difficulties motivate the study of the sharp-interface limit
$\eps\to0$ of the Cahn--Hilliard system. 
The limiting equations, the Mullins--Sekerka system
(see \eqref{MulSek} below), were formally derived by
Pego~\cite{Pego1989}. The convergence of the Cahn--Hilliard system
to the Mullins--Sekerka system was subsequently justified rigorously by
Alikakos--Bates--Chen~\cite{AlBaCh1994}, provided that the limiting
system admits a smooth solution. Local well-posedness and instantaneous
regularization of the limiting evolution were established by
Chen--Hong--Yi~\cite{ChHoYi1996}. 
For a broader discussion of related
sharp-interface limits, we refer the reader to the recent reviews of  
Abels and Garcke~\cite{AbGa2016} and Du and Feng~\cite{DuFe2020}. 

The mathematical theory justifying the sharp-interface limit breaks down
when diffuse transition layers overlap, thereby excluding topological
transitions such as coalescence, a fundamental feature of
phase coarsening.  Numerical solvers based on sharp-interface
formulations encounter the same difficulty; the method of
Zhu--Chen--Hou~\cite{ZhChHo1996}, for example, necessarily terminates
prior to a topology change.  Moreover, the authors explicitly note that
extensions to more complex phenomena and larger-scale problems first
require a robust method for resolving topological singularities.

In this work, we introduce a novel mathematical and numerical model,
based on a hybrid sharp--diffuse formulation, for continuing
the evolution of an interface through \emph{intermittent}
and \emph{isolated} topological transitions arising from Cahn--Hilliard dynamics.\footnote{%
To our knowledge, the underlying hybrid viewpoint was first suggested by
Lowengrub and Truskinovsky~\cite{LoTr1998}.} 
We formulate this approach in three principal steps:
\begin{itemize}
    \item approximation of the diffuse solution during ``regular'' evolution by
             its sharp-interface limit, together with an equivalent boundary-integral formulation;
    \item geometric characterization of the sharp-interface regime, with
             its breakdown triggering the suspension of the
             sharp-interface evolution;
    \item and, finally, resolution of the topological transition through a
             localized diffuse relaxation, followed by restart of the
             sharp-interface evolution.
\end{itemize}
Numerically, these three steps are realized using established
computational tools: the $\theta$--$L$ representation of the evolving
interface \cite{ZhChHo1996}, the manifold-death algorithm
\cite{ChMaPoZa2022}, and a smooth auxiliary variable (SAV) scheme
\cite{ShXuYa2018}, respectively. 

More generally, we note that many other numerical techniques 
have been developed to handle
topological changes in fluid flows, including Chorin's
vortex-filament reconnection algorithm \cite{Chorin1990}, the
front-tracking methods of Glimm et al.~\cite{Glimm2000} and
Tryggvason et al.~\cite{Tryggvason2001}, and specialized procedures for
boundary-integral simulations \cite{CrBlLo2001,ZiDa2013}.
To the best of our knowledge, however, all such methods resolve the
topology change by numerical regularization, often through an ad hoc rule.  
Lowengrub and Truskinovsky~\cite{LoTr1998} have demonstrated that the
interfacial stresses are sensitive to the choice of regularization when
geometric length scales become comparable to the interfacial thickness,
as occurs near a topology change.
Our sharp--diffuse model instead resolves the topological 
transition through physical diffuse-interface dynamics.

To illustrate numerically the proposed methodology, we consider a
four-circle coalescence problem originally studied by
Zhu--Chen--Hou~\cite{ZhChHo1996}.  Whereas their sharp-interface
calculation terminated before coalescence, our method resolves the
topology change and continues the evolution beyond it.  
Moreover, benchmarks against an established energy-stable
pseudospectral solver for the fully diffuse system show speedups of two
to three \emph{orders of magnitude}.  
We refer the reader to
\Cref{fig:four-circle-evolution,fig:four-circle-resolution-comparison} below
for a depiction of the complete evolution---from the breakdown of the 
sharp-interface regime, through localized diffuse relaxation, to the
restarted sharp-interface evolution---realized by our sharp--diffuse
numerical model.

\subsection*{Outline of the paper}
The remainder of the paper is structured as follows. 
\Cref{sec:diffuse-interface} introduces the diffuse-interface model and
recalls its basic properties.
\Cref{sec:sharp-limit} describes the basic geometry for the
formal passage to the sharp-interface limit, recalls the known
mathematical results justifying the sharp-interface approximation, and
presents an equivalent boundary-integral formulation of the limiting
evolution. The main contributions appear in
\Cref{sec:topological-events,sec:localized-relaxation,sec:surgery},
which formulate the sharp--diffuse continuation
procedure described above. We summarize our simple 
two-dimensional numerical implementation in
\Cref{sec:implementation} and test the method on the
four-circle coalescence problem in \Cref{sec:simulations}.
Finally, we provide concluding remarks in \Cref{sec:conclusions}.
Appendix~\ref{app:polynomial-profile} derives an explicit polynomial 
profile for idealized coalescence, and a video of the four-circle
simulation produced using our sharp--diffuse method is available at
\cite{RamaniWebsite}.


\section{Diffuse-interface model}
\label{sec:diffuse-interface}

\begin{figure}[tbhp]
\centering
\subfloat[Diffuse-interface geometry]{%
  \label{fig:phase-field-geometry}%
  \includegraphics[width=0.6\textwidth]{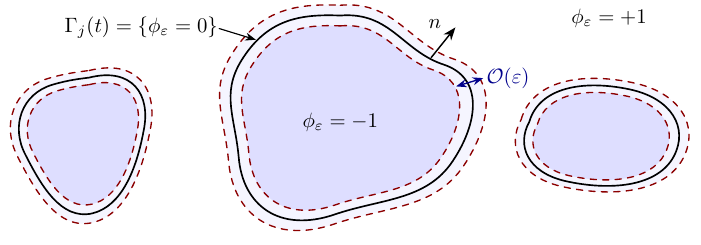}%
}
\hspace{2em}
\subfloat[Equilibrium profiles]{%
  \label{fig:equilibrium-profiles}%
  \includegraphics[width=0.3\textwidth]{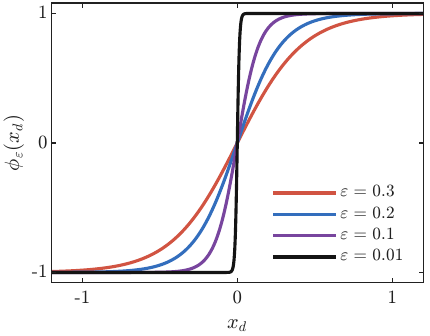}%
}
\caption{
\textbf{Left:} Schematic of the phase-field problem. The solid curves
represent the level sets $\phi_\eps=0$, while the dashed curves mark
transition layers of thickness $\O(\eps)$ separating the two
pure phases $\phi=\pm1$.
\textbf{Right:} Equilibrium profile $q(x_d/\eps)$, defined in
\eqref{transition-profile}, for several values of $\eps$.
}
\label{fig:schematic}
\end{figure}

Consider a binary mixture occupying $\R^d$, for $d=2,3$, with 
its local composition described by a 
\emph{phase-field function}
\begin{equation}\label{phase-field}
\phi_\eps : \R^d \times \R^+  \to [-1,1]. 
\end{equation}
The values $\pm1$ represent the two pure phases, which are assumed 
to be separated by a diffuse transition layer of thickness $\O(\eps)$. 
A schematic of the setup is shown in \Cref{fig:schematic}.
For each time $t > 0$, the mixture evolves according to
\begin{subequations}\label{CH}
\begin{align}
\p_t\phi_\eps (x,t)
&=m_\eps\Delta\mu_\eps (x,t),
\label{CH-phi}\\
\mu_\eps (x,t)
&=\sigmah 
\left(
-\eps\Delta\phi_\eps (x,t)
+\frac{1}{\eps}W'(\phi_\eps)
\right). 
\label{CH-mu}
\end{align}
\end{subequations}
Here $\mu_\eps(x,t)$ is the chemical potential, which measures the
first-order change in free energy produced by a local change in composition;
spatial gradients of $\mu_\eps$ drive diffusive transport, with the
mobility parameter $m_\eps>0$ controlling the rate of this transport.  
The constant $\sigmah >0$ is the diffuse capillary-energy coefficient.

The function $W$ is a smooth, nonnegative double-well potential satisfying
\begin{subequations}\label{double-well}
\begin{equation}\label{double-well-assumptions}
W(\pm1)=W'(\pm1)=0,
\qquad
W''(\pm1)>0.
\end{equation}
In this work, we consider the standard quartic polynomial approximation
\begin{equation}\label{double-well-eq}
W(\phi)=\tfrac14(1-\phi^2)^2.
\end{equation}
\end{subequations}

The system \eqref{CH} is initialized with
\begin{equation}\label{ICs}
\phi_\eps(x,0) = \mathring{\phi}_{\eps}(x), \qquad x\in\R^d, 
\end{equation}
where $\mathring{\phi}_{\eps}(x)$ is the prescribed initial datum. 
No initial condition is imposed on $\mu_\eps(x,0)$, which is
determined from $\phi_\eps(x,0)$ by \eqref{CH-mu}. 

We consider the equations \eqref{CH} posed on $\R^d$ 
with the far-field conditions
\begin{equation}\label{far-field}
\phi_\eps(x,t)\longrightarrow +1,
\quad
\mu_\eps(x,t)\longrightarrow 0
\qquad
\text{sufficiently rapidly as }|x|\longrightarrow\infty.
\end{equation}

\subsection{Free energy and chemical potential}

The system \eqref{CH} is the \emph{Cahn--Hilliard system}, derived by
Cahn and Hilliard~\cite{CaHi1958} to model phase separation and
coarsening in binary mixtures. In this model, the interfacial energy
carried by a finite-width transition layer is described by the
Ginzburg--Landau free energy, 
\begin{equation}\label{eng}
\Eng_\eps[\phi]
=
\sigmah 
\int_{\R^d}
\left(
\frac{\eps}{2}|\nabla\phi|^2
+\frac{1}{\eps}W(\phi)
\right)
\dd{x}. 
\end{equation}
The potential term causes the system to favor configurations containing the
two pure phases $\phi=\pm1$, while the gradient term penalizes rapid spatial
variation of the phase field. The two terms are weighted according to the
Modica--Mortola scaling \cite{MoMo1977}, producing diffuse
transition layers of thickness $\O(\eps)$ between the pure phases 
(see \Cref{lem:transition-profile} below).

The chemical potential describes the first-order change in free energy
produced by a local change in $\phi$, and thereby determines the
energetic driving force for diffusion.  
More precisely, for a smooth compactly supported test function
$\psi\in C_c^\infty(\R^d)$, the first variation of
\eqref{eng} in the direction $\psi$ is
\begin{equation*}
\begin{aligned}
D\Eng_\eps[\phi](\psi)
\coloneqq
\left.
\frac{\dd{\,}}{\dd{\delta}}
\Eng_\eps[\phi+\delta\psi]
\right|_{\delta=0}
&=
\sigmah 
\int_{\R^d}
\left(
-\eps\Delta\phi
+\frac{1}{\eps}W'(\phi)
\right)
\psi
\dd{x}, 
\end{aligned}
\end{equation*}
and since $\psi$ is arbitrary, it follows from \eqref{CH-mu} that
\begin{equation}\label{mu-var}
\mu_\eps
=
\frac{\delta\Eng_\eps}{\delta\phi}
[\phi_\eps]. 
\end{equation}

\subsection{Equilibrium transition profile}

The structure of a single flat diffuse interface can be determined directly
from the free energy.  Consider a stationary layer centered on the
plane $\{x_d=0\}$, and write
\begin{equation}\label{planar-ansatz}
\phi_\eps(x) =  q\left(\frac{x_d}{\eps}\right),
\qquad
\lim_{z\to-\infty}q(z)=-1,
\qquad
\lim_{z\to+\infty}q(z)=+1.
\end{equation}
At equilibrium, the first variation of the free energy vanishes, $\mu_\eps=0$.
Substitution of \eqref{planar-ansatz} into \eqref{CH-mu} then gives
the one-dimensional equilibrium equation
\begin{equation}\label{profile-eq}
-q''+W'(q)=0,
\qquad
q(-\infty)=-1,
\qquad
q(+\infty)=+1.
\end{equation}
If $q(z)$ solves \eqref{profile-eq}, then so does
$q(z-z_0)$ for any $z_0\in\R$.  We center the transition layer at
$z=0$ by imposing $q(0)=0$.  
The resulting profile and the associated interfacial energy may be computed explicitly. 

\begin{lemma}[Equilibrium transition profile]
\label{lem:transition-profile}
For the potential \eqref{double-well-eq}, the unique increasing solution of
\eqref{profile-eq} satisfying $q(0)=0$ is
\begin{equation}\label{transition-profile}
q(z)
=
\tanh\left(\frac{z}{\sqrt{2}}\right).
\end{equation}
The associated interfacial energy is
\begin{equation}\label{sharp-surface-tension}
\sigma
\coloneqq
\sigmah c_W,
\qquad
c_W
\coloneqq
\int_{-\infty}^{\infty}|q'(z)|^2\dd{z}
=
\int_{-1}^{1}\sqrt{2W(s)}\,\dd{s}
=
\frac{2\sqrt{2}}{3}.
\end{equation}
\end{lemma}

Plots of the profile \eqref{transition-profile} 
are shown in \cref{fig:equilibrium-profiles} for different values of $\eps$.

\subsection{Diffuse transition-layer thickness}

The equilibrium profile \eqref{transition-profile} provides a natural
geometric description of the diffuse interface.
Let $\Gamma\subset\R^d$ be a
smooth closed interface enclosing a bounded region $D$, and define its
signed distance by
\begin{equation}\label{sdf-def}
d_\Gamma(x)
\coloneqq
\begin{cases}
-\dist(x,\Gamma),
&x\in D,
\\
\dist(x,\Gamma),
&x\in\R^d\setminus\overline D,
\\
0,
&x\in\Gamma.
\end{cases}
\end{equation}
The diffuse field associated with $\Gamma$ is obtained by placing the
one-dimensional profile \eqref{transition-profile} across the interface:
\begin{equation}\label{sdf-phi}
\phi_\eps^\Gamma(x)
\coloneqq
q\left(\frac{d_\Gamma(x)}{\eps}\right)
=
\tanh\left(
\frac{d_\Gamma(x)}{\eps\sqrt{2}}
\right).
\end{equation}

Since \eqref{transition-profile} 
approaches $\pm1$ only asymptotically, a finite
transition-layer width must be specified by fixing a pair of level sets.
For example, we can define an effective transition layer as the region
between the $\pm 0.9$ level sets of $\phi_\eps^\Gamma(x)$.
Using \eqref{sdf-phi}, this region can be written equivalently in
terms of $d_\Gamma(x)$ as
\begin{subequations}\label{transition-layer}
\begin{gather}
\Tlyr_\eps[\Gamma]
\coloneqq
\left\{
x\in\R^d:
\left|\phi_\eps^\Gamma(x)\right|<0.9
\right\}
=
\left\{
x\in\R^d:
|d_\Gamma(x)|<\Clyr\eps
\right\}, \\
\Clyr
=
\sqrt{2}\operatorname{arctanh}(0.9)
\approx
2, 
\label{Clayer}
\end{gather}
\end{subequations}
so that $\Clyr\eps \approx 2\eps$ is the effective half-width of the diffuse layer.

\subsection{Energy dissipation and mass conservation}

Equation \eqref{CH-phi} may be written in flux form with
diffusive flux $f_\eps$,
\begin{subequations}\label{flux-form}
\begin{align}
\p_t\phi_\eps (x,t)
+\nabla\cdot f_\eps (x,t)
&=0,
\label{flux-balance}
\\
f_\eps (x,t)
&=-m_\eps\nabla\mu_\eps (x,t).
\label{diffusive-flux}
\end{align}
\end{subequations}
The first equation expresses local conservation of composition, while the
second states that the diffusive flux is directed down gradients of the
chemical potential.  For the closed-interface configurations considered
here, we take $\phi_\eps=+1$ to be the exterior phase.  
The mass of the $-1$ phase is then defined by
\begin{equation}\label{phase-mass}
\Mass_\eps(t)
\coloneqq
\int_{\R^d}
\tfrac{1}{2} \bigl(1-\phi_\eps(x,t)\bigr)
\dd{x},
\end{equation}
provided $1-\phi_\eps(\cdot,t)\in L^1(\R^d)$. 

\begin{lemma}[Energy dissipation and mass conservation]
\label{lem:energy-mass}
Let $(\phi_\eps,\mu_\eps)$ be a sufficiently regular solution of
\eqref{CH} satisfying the stated decay assumptions, and suppose that
$1-\phi_\eps(\cdot,t)\in L^1(\R^d)$.  Then
\begin{equation}\label{energy-law}
\frac{\dd{\,}}{\dd{t}}
\Eng_\eps[\phi_\eps]
=
-m_\eps
\int_{\R^d}
|\nabla\mu_\eps|^2
\dd{x}
\leq0.
\end{equation}
Moreover, the phase mass \eqref{phase-mass} is conserved.
\end{lemma}

\subsection{The \texorpdfstring{$H^{-1}$}{H-1} gradient-flow structure}

The first variation of the free energy identifies
the chemical potential, but it does not by itself
determine the evolution.  The evolution is obtained only
after specifying the metric in which the free energy
decreases most rapidly. Because the gradient flow is
restricted to phase fields with fixed phase mass, every
admissible variation $\psi$ must satisfy $\int_{\R^d}\psi\dd{x}=0$.
We denote the homogeneous $H^{-1}$ inner product on smooth,
decaying, zero-mean functions by
\begin{equation}\label{H1-inner-product}
\langle u,v\rangle_{H^{-1}}
\coloneqq
\int_{\R^d}
(-\Delta)^{-1}u\,v
\dd{x}.
\end{equation}

\begin{lemma}[$H^{-1}$-gradient-flow structure]
\label{lem:gradient-flow}
Let $\phi_\eps$ be a smooth phase field satisfying
the stated decay assumptions, and let $\mu_\eps$ be
the chemical potential defined by \eqref{mu-var}.  Then
$\gradd_{H^{-1}}\Eng_\eps[\phi_\eps] = -\Delta\mu_\eps$, so that
\eqref{CH-phi} may be written as
\begin{subequations}
\begin{equation}\label{Hminus-one-flow}
\p_t\phi_\eps
=
-m_\eps
\gradd_{H^{-1}}\Eng_\eps[\phi_\eps],
\end{equation}
and the corresponding gradient-flow dissipation identity is
\begin{equation}\label{dissipation-identity}
\frac{\dd{\,}}{\dd t}
\Eng_\eps[\phi_\eps]
=
-m_\eps
\left\|
\gradd_{H^{-1}}\Eng_\eps[\phi_\eps]
\right\|_{H^{-1}}^2.
\end{equation}
\end{subequations}
\end{lemma}


\section{Sharp-interface limit}
\label{sec:sharp-limit}

We are interested in the sharp-interface limit of \eqref{CH} as
$\eps\to0$, corresponding to the physical regime in which the width
of the diffuse transition layer is negligible relative to the geometric
length scales of the phase domains.  We assume that the mobility remains
nondegenerate in this limit:
\begin{equation}\label{mobility-scaling}
m_\eps
=
m_0+\O(\eps),
\qquad
m_0>0,
\end{equation}
so that diffusion remains active in the bulk phases and disconnected
components can exchange mass through the surrounding mixture.

\subsection{Sharp-interface geometry}

Fix $T>0$ and consider $N$ pairwise disjoint closed interfaces
$\Gamma_j(t)$ evolving for $t\in[0,T]$, with $D^-_j(t)$ denoting the bounded
domain enclosed by $\Gamma_j(t)$. We write
\begin{subequations}\label{geometry}
\begin{equation}\label{phase-geometry}
\Gammabf(t)
\coloneqq
\bigcup_{j=1}^{N}\Gamma_j(t), 
\qquad
D^-(t)
\coloneqq
\bigcup_{j=1}^{N}D^-_j(t), 
\qquad 
D^+(t) \coloneqq \R^d\setminus\overline{D^-(t)}
\end{equation}
We assume that each interface component admits a parametrization
\begin{equation}\label{parametrization}
X_j
:
\S^{d-1}\times[0,T]
\longrightarrow
\R^d,
\qquad
\Gamma_j(t)
=
X_j(\S^{d-1},t),
\end{equation}
where $X_j(\cdot,t)$ is an embedding for every $t\in[0,T]$ and
\begin{equation}\label{interface-regularity}
X_j
\in
C^\infty\bigl(
\S^{d-1}\times[0,T];
\R^d
\bigr).
\end{equation}
\end{subequations}
Thus, each $\Gamma_j(t)$ is a smooth closed hypersurface.

The total interfacial measure is denoted by
\begin{equation}\label{interface-area}
|\Gammabf(t)|
\coloneqq
\mathcal{H}^{d-1}\bigl(\Gammabf(t)\bigr)
=
\sum_{j=1}^N
\mathcal{H}^{d-1}\bigl(\Gamma_j(t)\bigr),
\end{equation}
where $\mathcal{H}^{d-1}$ denotes the $(d-1)$-dimensional Hausdorff
measure. Thus, \eqref{interface-area} gives the total interfacial length
when $d=2$ and the total interfacial area when $d=3$.

The limiting phase field $\phi_0(x,t)$ is assumed to take the form 
\begin{equation}\label{phi-limit}
\phi_0(x,t)
=
\begin{cases}
-1,
&x\in D^-(t),
\\
+1,
&x\in D^+(t).
\end{cases}
\end{equation}
Equivalently,
$\phi_0(\cdot,t)=1-2\chi_{D^-(t)}$, where $\chi_A$ 
denotes the indicator function on the set $A$. 
With a slight abuse of notation, we denote the 
time-dependent signed-distance function and transition layers
using the same notation in \eqref{sdf-def} and \eqref{transition-layer}, 
\begin{equation*}
d_{\Gammabf}(x,t)
\coloneqq
d_{\Gammabf(t)}(x),
\qquad
\Tlyr_\eps(t)
\coloneqq
\Tlyr_\eps[\Gammabf(t)]. 
\end{equation*}

\subsubsection{Reach and signed-distance coordinates}

The geometry of $\Gammabf(t)$ can be described using 
smooth signed-distance (or normal) coordinates wherever each nearby
point has a unique nearest point on the interface.  The size of this
neighborhood is quantified by the following definition.

\begin{definition}[Reach \cite{Federer1969}]
\label{def:reach}
The \emph{reach} of a closed set $A\subset\R^d$ is defined as
\begin{equation}\label{reach}
\reach(A)
\coloneqq
\sup\left\{
\rho>0:
\begin{array}{c}
\text{every $x\in\R^d$ satisfying
$\dist(x,A)<\rho$}\\
\text{has a unique nearest point in $A$}
\end{array}
\right\}.
\end{equation}
\end{definition}

\begin{figure}[tbhp]
\centering
\includegraphics[width=0.6\textwidth]{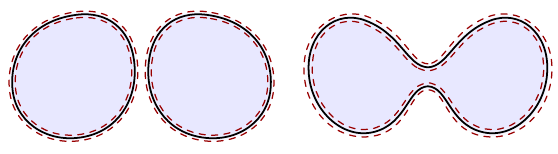}
\caption{
\textbf{Configuration satisfying the reach condition
\eqref{separated-layers}.}
The solid black curves represent the sharp interfaces, while the dashed
red curves delimit diffuse transition layers of effective half-width
$\Clyr\eps$. Although the reach is small, the transition layers remain
disjoint in both configurations. A further decrease in reach would cause
the layers to overlap, signaling an impending coalescence of distinct
components (left) or pinch-off of a single component (right).
}
\label{fig:reach}
\end{figure}

A minimal geometric requirement for the sharp-interface
approximation to remain valid is that the associated diffuse transition
layers do not overlap; see \Cref{fig:reach}. The
separation condition can be characterized directly in terms of the reach,
leading to the following basic assumption on the sharp-interface evolution.

\begin{assumption}[Separated transition layers]
\label{ass:separated-layers}
The transition layers remain well separated on $[0,T]$, in the sense that
\begin{equation}\label{separated-layers}
\reach\bigl(\Gammabf(t)\bigr)
>
\Clyr\eps
\qquad
\text{for every }t\in[0,T],
\end{equation}
where $\Clyr$ is defined in \eqref{transition-layer}.
\end{assumption}

On any interval $[0,T]$ on which \Cref{ass:separated-layers} holds, define
\begin{equation}\label{reach-inf}
\rho_*
\coloneqq
\inf_{t\in[0,T]}
\reach\bigl(\Gammabf(t)\bigr)
\ge \Clyr \eps.
\end{equation}
For any $0<\rho<\rho_*$, we define a tubular neighborhood of
$\Gammabf(t)$ by
\begin{equation}\label{tube}
\mathcal{N}_\rho(t)
\coloneqq
\left\{
x\in\R^d:
\dist\bigl(x,\Gammabf(t)\bigr)<\rho
\right\},
\qquad
\mathcal{N}_\rho
\coloneqq
\bigcup_{0\leq t\leq T}
\mathcal{N}_\rho(t)\times\{t\}, 
\end{equation}
where $\mathcal{N}_\rho\subset\R^d\times[0,T]$ denotes the spacetime tube
swept out by $\mathcal{N}_\rho(t)$.

The assumed smoothness of the parametrization \eqref{interface-regularity} 
and the uniform reach bound \eqref{separated-layers} yield the following 
normal-coordinate representation.\footnote{The
fixed-time statement follows from \cite[Lemma~2.8]{DzEl2013}, while the
smooth dependence on time follows from \eqref{interface-regularity}.}

\begin{lemma}[Signed-distance coordinates]
\label{lem:sdf-regularity}
For every $0<\rho<\rho_*$ and every
$(x,t)\in\mathcal{N}_\rho$, there is a unique nearest point
$\pi(x,t)\in\Gammabf(t)$. Moreover,
\begin{equation}\label{sdf-regularity}
d_{\Gammabf}
\in
C^\infty(\mathcal{N}_\rho),
\qquad
\pi
\in
C^\infty(\mathcal{N}_\rho;\R^d),
\end{equation}
and each $x\in\mathcal{N}_\rho(t)$ admits the
representation
\begin{equation}\label{normal-coords}
x = \pi(x,t) +  d_{\Gammabf}(x,t)\nabla d_{\Gammabf}(x,t),
\qquad
|\nabla d_{\Gammabf}(x,t)|
=
1.
\end{equation}
\end{lemma}

\subsubsection{Normal and curvature conventions}

Let $\alpha=(\alpha_1,\ldots,\alpha_{d-1})$ denote local coordinates on
$\S^{d-1}$. Because $X_j(\cdot,t)$ is a smooth embedding, we may define 
a smooth unit normal on $\Gamma_j(t)$ by
\begin{subequations}\label{normal}
\begin{equation}\label{normal-interface}
n\bigl(X_j(\alpha,t),t\bigr)
\coloneqq
\begin{cases}
\displaystyle
\frac{\bigl(\p_{\alpha}X_j\bigr)^\perp}
     {|\p_{\alpha}X_j|},
& d=2,
\\[2ex]
\displaystyle
\frac{\p_{\alpha_1}X_j\times\p_{\alpha_2}X_j}
     {|\p_{\alpha_1}X_j\times\p_{\alpha_2}X_j|},
& d=3,
\end{cases}
\end{equation}
where $v^\perp=(-v_2,v_1)$ and the local coordinates are oriented so that
$n$ points from the $-1$ phase into the $+1$ phase.
By \cref{lem:sdf-regularity}, the signed-distance function is smooth
throughout $\mathcal{N}_\rho$, and therefore defines the canonical
extension of the interface normal
\begin{equation}\label{normal-extension}
n(x,t)
\coloneqq
\nabla d_{\Gammabf}(x,t)
=
n\bigl(\pi(x,t),t\bigr),
\qquad
(x,t)\in\mathcal{N}_\rho.
\end{equation}
It then follows from \eqref{normal-coords} that every point in 
$\mathcal{N}_{\rho}$ can be represented as 
\begin{equation}
x = \pi (x,t) + d_{\Gammabf}(x,t) \,n (x,t). 
\end{equation}
\end{subequations}

For $(x,t)\in\mathcal{N}_\rho$, define the tangential projection
$P_\tau \coloneqq I-n\otimes n$. 
At each $x\in\mathcal{N}_\rho(t)$, $P_\tau$ projects orthogonally onto
the tangent space of the signed-distance level set through $x$; when
$x\in\Gammabf(t)$, this space is precisely $T_x\Gammabf(t)$. 
For a scalar function $f$ and a vector field $v$, we
define the normal derivative, tangential gradient, and tangential
divergence by
\begin{equation*}
\p_n f
\coloneqq
n\cdot\nabla f,
\qquad
\p_\tau f
\coloneqq
P_\tau\nabla f,
\qquad
\p_\tau\cdot v
\coloneqq
\trace(P_\tau\nabla v).
\end{equation*}

The scalar mean curvature is defined by
\begin{subequations}\label{kappa}
\begin{equation}
\kappa
\coloneqq
\p_\tau\cdot n
=
\trace(P_\tau\nabla n)
\qquad
\text{on }\Gammabf(t).
\end{equation}
With this convention, a sphere of radius $R$ has curvature
$\kappa=(d-1)/R>0$. The extension of the curvature off the interface
is given by 
\begin{equation}\label{kappa-sdf}
\kappa
=
\Delta d_{\Gammabf}
\qquad
\text{for } (x,t) \in \mathcal{N}_\rho.
\end{equation}
\end{subequations}
It follows from \eqref{sdf-regularity} that
\begin{equation}\label{CH-geometric-field-regularity}
n
\in
C^\infty(\mathcal{N}_\rho;\R^d),
\qquad
\kappa
\in
C^\infty(\mathcal{N}_\rho).
\end{equation}

\subsubsection{Normal velocity and jump notation}

The normal velocity of $\Gamma_j(t)$ is denoted by 
\begin{equation}\label{V-normal}
V_j(\alpha,t)
\coloneqq
\p_tX_j(\alpha,t)\cdot
n(X_j(\alpha,t),t),
\end{equation}
so that $V_j>0$ corresponds to motion into the exterior phase.  The
signed-distance regularity
\eqref{sdf-regularity} allows us to differentiate
$d_{\Gammabf}(X_j(\alpha,t),t)=0$ to obtain
\begin{equation}\label{sdf-velocity}
\p_td_{\Gammabf}
\bigl(X_j(\alpha,t),t\bigr)
=
-V_j(\alpha,t).
\end{equation}

Finally, if a quantity $g$ has traces on both sides of the interface, we
write
\begin{equation}\label{jump}
g^\pm(X,t)
\coloneqq
\lim_{h\downarrow0}g(X\pm hn(X,t),t),
\qquad
[g]
\coloneqq
g^+-g^-.
\end{equation}

\subsection{The Mullins--Sekerka limit system}

Under the regularity and separation hypotheses above, the signed-distance
coordinates furnished by \Cref{lem:sdf-regularity} provide the geometric
framework for a formal passage to the limit $\eps\to0$ in the
Cahn--Hilliard system \eqref{CH}. 
The Modica--Mortola theorem \cite{MoMo1977,Modica1987} identifies the
limiting diffuse energy with $\sigma|\Gammabf(t)|$, where
$|\Gammabf(t)|$ is the total interfacial measure \eqref{interface-area} 
and $\sigma$ is the interfacial energy 
\eqref{sharp-surface-tension} associated with the equilibrium profile $q$.

Using matched asymptotic expansions, Pego~\cite{Pego1989} showed 
that the chemical potential admits the formal leading-order limit in
each bulk phase, 
\begin{equation}\label{chemical-potential-limit}
\lim_{\eps\to 0}
\mu_\eps(x,t)
\
= 
\
\widehat{\mu}_0^\pm(x,t), 
\qquad
\text{locally in }D^\pm(t), 
\end{equation}
where $\widehat{\mu}_0$ is governed by the 
Mullins--Sekerka system \eqref{MulSek} below. 
In the bulk, the leading phase field takes the constant values
$\phi_0^\pm=\pm1$, so that its time derivative vanishes and the
leading-order mass-conservation equation reduces to
$\Delta\widehat{\mu}_0^\pm=0$. Near the interface, the signed-distance
coordinate in \eqref{normal-coords} is rescaled by setting
$z=d_{\Gammabf}(x,t)/\eps$. In these inner coordinates, the leading-order
phase field is the equilibrium profile $q(z)$ in
\eqref{transition-profile}, which connects the two bulk values.
Matching the inner and outer expansions yields continuity of
$\widehat{\mu}_0$ across the interface, with the trace given by
the so-called Gibbs--Thomson condition. 
The inner mass balance gives the normal velocity in terms of
$[\p_n\widehat{\mu}_0]$, so that the jump in the normal derivative drives
the interface motion.

At leading order, conservation of the diffuse phase mass reduces to
conservation of the total volume of $D^-(t)$. Enforcing this global
constraint introduces a time-dependent scalar Lagrange multiplier,
which we denote by $\mu^{\mathrm f}(t)$, into the Gibbs--Thomson
condition:
\begin{equation}\label{shifted-Gibbs-Thomson}
\widehat{\mu}_0^-
=
\widehat{\mu}_0^+
=
-\frac{\sigma}{2}\kappa-\mu^{\mathrm f}(t)
\qquad
\text{on }\Gammabf(t).
\end{equation}
It is convenient to absorb this scalar into the limiting chemical potential
by defining
\begin{equation}\label{sharp-chemical-potential-shift}
\mu_0^\pm
\coloneqq
\widehat{\mu}_0^\pm+\mu^{\mathrm f}(t), 
\end{equation}
which yields the standard Gibbs--Thomson condition \eqref{GibbsThomson}. 
Moreover, \eqref{far-field} becomes the far-field condition \eqref{MS-far-field}. 
The resulting sharp-interface system
is summarized as follows.

\begin{theorem}[Formal sharp-interface limit \cite{Pego1989}]
\label{thm:sharp-limit}
Let $(\phi_\eps,\mu_\eps)$ be a family of smooth solutions of
\eqref{CH}. Assume the mobility scaling \eqref{mobility-scaling}, 
matched inner and outer expansions about $\Gammabf(t)$, and
that the smooth interfaces satisfy 
\Cref{ass:separated-layers} throughout $[0,T]$. Then, formally at leading
order as $\eps\to0$, $\phi_\eps\to\phi_0$, where $\phi_0$ is given by
\eqref{phi-limit}, and
$\mu_\eps\to\mu_0^\pm-\mu^{\mathrm f}(t)$ locally in $D^\pm(t)$, where
\begin{subequations}\label{MulSek}
\begin{align}
\Delta\mu_0^\pm
&=0
&&\text{in }D^\pm(t),
\label{MS-bulk}
\\
\mu_0^-
=
\mu_0^+
&=
-\frac{\sigma}{2}\kappa
&&\text{on }\Gammabf(t),
\label{GibbsThomson}
\\
V_j
&=
-\frac{m_0}{2}[\p_n\mu_0]
&&\text{on }\Gamma_j(t),
\qquad j=1,\ldots,N,
\label{MS-velocity}
\\
\mu_0^+(x,t)
&\longrightarrow
\mu^{\mathrm f}(t)
&&\text{as }|x|\longrightarrow\infty.
\label{MS-far-field}
\end{align}
\end{subequations}
Here $\sigma$ is defined in \eqref{sharp-surface-tension}, and
$\mu^{\mathrm f}\in C^\infty([0,T])$ is the far-field Lagrange multiplier
enforcing global mass conservation.
\end{theorem}

For well-prepared initial data, Alikakos--Bates--Chen~\cite{AlBaCh1994} 
rigorously justify \cref{thm:sharp-limit} on any
fixed time interval over which the Mullins--Sekerka evolution \eqref{MulSek}
admits a smooth classical solution.
The existence and uniqueness of such solutions 
are given by the following result due to Chen--Hong--Yi~\cite{ChHoYi1996}.\footnote{This
is a slight restatement of Theorem~1.1 in
\cite{ChHoYi1996}, which is formulated on a
bounded domain with homogeneous Neumann boundary conditions.  
We use the corresponding whole-space formulation.}

\begin{theorem}[Strong Mullins--Sekerka solutions \cite{ChHoYi1996}]
\label{thm:CHY}
Let $r\in(0,1)$, and suppose the initial interface
$\Gammabf(0)$ is a compact, embedded hypersurface satisfying
\begin{equation}\label{CHY-initial-regularity}
\Gammabf(0)
\in
C^{3+r}.
\end{equation}
Then there exists $T>0$ such that
\eqref{MulSek} admits a unique classical solution
on $[0,T]$, with the space--time interface belonging to the class
\begin{subequations}
\begin{equation}\label{CHY-regularity}
\bigcup_{0\leq t\leq T}
\bigl(\Gammabf(t)\times\{t\}\bigr)
\in
C^{3+r,(3+r)/3}.
\end{equation}
Moreover, for every $s\in(0,T)$,
\begin{equation}\label{CHY-smoothing}
\bigcup_{s\leq t\leq T}
\bigl(\Gammabf(t)\times\{t\}\bigr)
\in
C^\infty.
\end{equation}
\end{subequations}
Here $C^r$ and $C^{r,r/3}$ denote the spatial
and anisotropic space--time H\"older spaces, respectively 
(see \cite{ChHoYi1996} for precise definitions).
\end{theorem}

\begin{remark}
The instantaneous smoothing property
\eqref{CHY-smoothing} ensures that, for every $s\in(0,T)$,
the evolving interfaces admit parametrizations satisfying
\eqref{interface-regularity} on $[s,T]$, thereby justifying the
smoothness assumed in the matched-asymptotic expansion.
\end{remark}

\begin{remark}
The conservation and gradient-flow structure
of the diffuse system (cf. \Cref{lem:energy-mass})  pass
formally to the sharp-interface limit as conservation of the enclosed
volume and dissipation of the total interfacial area.
\end{remark}

\subsection{Boundary integral formulation}

At each fixed time, the bulk harmonic problem in
\eqref{MulSek} can be eliminated in favor of an
integral equation posed only on the moving interface using 
a single-layer potential representation. 

\begin{proposition}[Boundary-integral formulation \cite{Chen1993,ZhChHo1996}]
\label{prop:BI}
Let $\Gammabf(t)$ be a family of
smooth disjoint closed interfaces. 
We denote the fundamental solution of
$-\Delta$ by
\begin{equation}\label{Gker}
\Gker(x)
=
\begin{cases}
\displaystyle
-\frac{1}{2\pi}\log|x|,
&d=2,
\\[0.8em]
\displaystyle
\frac{1}{4\pi|x|},
&d=3,
\end{cases}
\qquad x\neq0.
\end{equation}
For a smooth density $\lambda$ on $\Gammabf(t)$, define the
single-layer potential
\begin{equation}\label{SL-pot}
\Sop_{\Gammabf(t)}[\lambda](x)
\coloneqq
\int_{\Gammabf(t)}
\Gker(x-y)\lambda(y)\dd S_y.
\end{equation}
Let $V$ denote the normal velocity on $\Gammabf(t)$, defined by
\[
V\bigl(X_j(\alpha,t),t\bigr)
\coloneqq
V_j(\alpha,t).
\]
Then the Mullins--Sekerka system \eqref{MulSek} is equivalent to the
coupled boundary-integral system
\begin{subequations}\label{BI-eqs}
\begin{align}
\mu^{\mathrm f}(t)
+
\frac{2}{m_0}
\Sop_{\Gammabf(t)}[V](x)
&=
-\frac{\sigma}{2}\kappa(x,t),
&&x\in\Gammabf(t),
\label{BI-eq}
\\
\int_{\Gammabf(t)}
V(y,t)\dd S_y
&=0.
\label{BI-constraint}
\end{align}
\end{subequations}
For a given interface configuration, \eqref{BI-eqs} determines
$V$ and $\mu^{\mathrm f}(t)$ simultaneously. The chemical potential is
recovered from
\begin{equation}\label{mu-SL}
\mu_0(x,t)
=
\mu^{\mathrm f}(t)
+
\frac{2}{m_0}
\Sop_{\Gammabf(t)}[V](x),
\qquad
x\in\R^d\setminus\Gammabf(t),
\end{equation}
and the interface parametrizations evolve according to
\begin{equation}\label{BI-interface}
\p_tX_j(\alpha,t)
=
V_j(\alpha,t) \,n\bigl(X_j(\alpha,t),t\bigr)
+
V_j^\tau(\alpha,t),
\qquad
V_j^\tau(\alpha,t)
\in
T_{X_j(\alpha,t)}\Gamma_j(t),
\end{equation}
for $j=1,\ldots,N$. The tangential velocities $V_j^\tau$ may be chosen freely. 
\end{proposition}

\begin{proof}
We prove the equivalence at each fixed time. For the forward implication,
suppose that $(\Gammabf,\mu_0)$ is a smooth solution of
\eqref{MulSek}. Since $\mu_0$ is continuous across $\Gammabf(t)$ and harmonic in the
two bulk phases, its distributional Laplacian is
\begin{equation}\label{distributional-laplacian}
\Delta\mu_0
=
[\p_n\mu_0]\,\delta_{\Gammabf(t)},
\qquad
\left\langle
\delta_{\Gammabf(t)},\psi
\right\rangle
\coloneqq
\int_{\Gammabf(t)}
\psi\dd S.
\end{equation}
The velocity condition \eqref{MS-velocity} therefore gives
\begin{equation}\label{distributional-Poisson-equation}
-\Delta\mu_0
=
\frac{2}{m_0}
V\,\delta_{\Gammabf(t)}.
\end{equation}
The single-layer potential \eqref{SL-pot} is harmonic away from the
interface and continuous across it, while its normal derivative satisfies
the standard jump relation
\begin{equation}\label{SL-jump}
\left[
\p_n
\Sop_{\Gammabf(t)}[\lambda]
\right]
=
-\lambda.
\end{equation}
Consequently, the solution of
\eqref{distributional-Poisson-equation} satisfying the far-field
condition \eqref{MS-far-field} is given by \eqref{mu-SL}. Taking its
trace on $\Gammabf(t)$ and imposing the Gibbs--Thomson condition
\eqref{GibbsThomson} yields \eqref{BI-eq}.
Global mass conservation requires the total volume of $D^-(t)$ to
remain constant. The transport identity therefore gives
\begin{equation*}
0
=
\frac{\dd{\,}}{\dd t}|D^-(t)|
=
\int_{\Gammabf(t)}
V(y,t)\dd S_y,
\end{equation*}
which is precisely \eqref{BI-constraint}. In the coupled system
\eqref{BI-eqs}, the scalar $\mu^{\mathrm f}(t)$ is the Lagrange
multiplier selected simultaneously with $V$ so that this constraint is
satisfied. Finally, decomposing
$\p_tX_j$ into its normal and tangential components gives
\eqref{BI-interface}.

For the converse implication, suppose that $V$ and
$\mu^{\mathrm f}(t)$ satisfy \eqref{BI-eqs} and that the interface
parametrizations evolve according to \eqref{BI-interface}. Define
$\mu_0$ by \eqref{mu-SL}. The single-layer potential is harmonic away
from $\Gammabf(t)$, so \eqref{MS-bulk} holds, and its continuity across
the interface, together with \eqref{BI-eq}, gives the Gibbs--Thomson
condition \eqref{GibbsThomson}. Moreover, the jump relation
\eqref{SL-jump} gives the velocity condition \eqref{MS-velocity}.
In three dimensions, the constraint \eqref{BI-constraint} removes the leading monopole term
in the expansion of \eqref{Gker}, whereas in two dimensions it removes
the leading logarithmic term. Consequently, 
$\Sop_{\Gammabf(t)}[V](x)=\O(|x|^{-2})$ when $d=3$ and
$\Sop_{\Gammabf(t)}[V](x)=\O(|x|^{-1})$
when $d=2$, as $|x|\to\infty$.  Hence
\eqref{MS-far-field} also holds, completing the equivalence.
\end{proof}

\begin{remark}
The equivalence in \Cref{prop:BI} is understood at the level of smooth
solutions. For the forward implication, the required regularity is
provided by \Cref{thm:CHY}. For the converse implication, Chen~\cite{Chen1993} 
has established local-in-time existence for the direct boundary-integral evolution of a
single interface in two dimensions; to the best of our knowledge,
uniqueness and regularity of such solutions, along with the three-dimensional case, remain open.
\end{remark}

\begin{remark}
In two dimensions, the tangential velocity in \eqref{BI-interface} may
be chosen to preserve equal-arclength parametrizations in the so-called 
$\theta$-$L$ formulation \cite{HoLoSh1994}. 
The boundary-integral equations in this formulation can be solved accurately and efficiently
by pairing Fourier spectral discretization in space with semi-implicit
time integration of the stiff leading-order terms, 
see \cite{HoLoSh1994,ZhChHo1996} and \Cref{sec:implementation}.
\end{remark}


\section{Intermittent and isolated topological transitions}
\label{sec:topological-events}

To summarize, two PDE systems have been introduced thus far:
the Cahn--Hilliard system \eqref{CH}
and the Mullins--Sekerka system \eqref{MulSek}.
Under \Cref{ass:separated-layers},
\Cref{thm:sharp-limit} identifies the latter
as the formal sharp-interface limit of the former.
For sufficiently regular initial interfaces,
\Cref{thm:CHY} gives a unique local
classical solution of this limit system, and 
\Cref{prop:BI} gives its equivalent
boundary integral formulation, for which efficient
and accurate numerical solvers are available.
Taken together, these results justify the classical
sharp-interface evolution while the transition layers 
remain separated.
Consequently, they do not permit topological events
such as coalescence.

The Alikakos--Bates--Chen result \cite{AlBaCh1994} is
restricted to precisely this classical regime, since it assumes 
the existence of a uniform tubular neighborhood on
which the signed-distance function is smooth. When
\Cref{ass:separated-layers} fails, the medial axis enters the transition
layer, the closest-point projection ceases to be unique, and the
signed-distance coordinates underlying the matched-asymptotic
construction break down.

\begin{figure}[tbhp]
\centering
\includegraphics[width=0.6\textwidth]{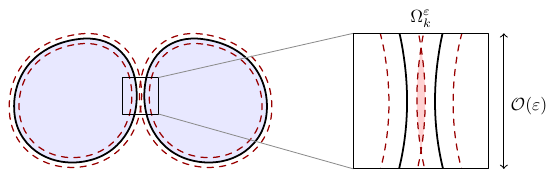}
\caption{
\textbf{Localized failure of the reach condition
\eqref{separated-layers}.}
The solid black curves represent two approaching sharp interfaces, while
the dashed red curves delimit their diffuse transition layers. The inset
magnifies an $\O(\eps)$ neighborhood $\Omega_k^\eps$ in which
the layers overlap and the closest-point projection becomes nonunique
along the medial axis.
}
\label{fig:reach-fail}
\end{figure}

\subsection{Weak continuation through varifold solutions}

One way to interpret the continuation of the sharp-interface evolution
beyond a topological singularity is to adopt a weaker solution concept.
This is the approach taken by Chen~\cite{Chen1996}, who proved that
a subsequence of Cahn--Hilliard solutions converges as
$\eps\to0$ to a global varifold solution of the Mullins--Sekerka system, which
remains meaningful without a smooth embedded interface. Such a weak
(measure-valued) continuation does not, however, provide a unique smooth interface from
which to restart the classical evolution.

\subsection{Strong continuation through interface surgery}

An alternative program, outlined by Lowengrub and
Truskinovsky~\cite{LoTr1998}, seeks to resolve topological transitions
through diffuse mixing dynamics localized in both space and time.
While the transition layers remain
well separated, their internal structure may be neglected and the
sharp-interface system governs the evolution. When two layers
overlap, however, the diffuse mixing dynamics must be resolved. These
dynamics provide a smooth continuation through the topological
singularity, after which a new separated sharp interface
emerges. In this picture, the regularization and its associated energy
dissipation are localized in space and time near the topological event.

The strong continuation of the solution is realized here through a 
formal \emph{interface surgery} algorithm, which will be described below. 
At a topological event time $T$, its purpose
is to take the incoming interface family
immediately before the event $t = T^-$ and return the outgoing
interface family immediately after the event $t = T^+$: 
\begin{equation}\label{interface-families}
\Gammabf(T^\pm)
=
\bigcup_{j=1}^{N(T^\pm)}
\Gamma_j(T^\pm),
\qquad
\Gamma_j(T^\pm)
=
X_j(\S^{d-1},T^\pm). 
\end{equation}
Note that the component counts $N(T^-)$ and $N(T^+)$
need not agree.

\subsection{Topological events}

Let $r\in(0,1)$, and let
$\Gammabf(0)$ be a compact, embedded
$C^{3+r}$ hypersurface satisfying the reach condition \eqref{separated-layers}. 
By \Cref{thm:CHY},
the initial interface generates a unique local
classical solution $\Gammabf(t)$ of
the Mullins--Sekerka system.
On any interval where
\Cref{ass:separated-layers} holds,
\Cref{thm:sharp-limit} justifies using
this solution as the leading-order approximation
to the diffuse-interface evolution.
Finally, \Cref{prop:BI} supplies
the equivalent boundary integral formulation used
to compute this classical solution.
Let us consider the evolution on
the time interval
\begin{equation*}
0 \le t \le \tmax,  \qquad \tmax > 0. 
\end{equation*}

\begin{definition}[Topological event times]
\label{def:event-times}
The first topological event time is
\begin{subequations}\label{event-times}
\begin{equation}\label{event-time-first}
T_0
\coloneqq
\inf
\left\{
t\in(0,\tmax]:
\reach\bigl(\Gammabf(t)\bigr)
\leq
\Clyr\eps
\right\}. 
\end{equation}
If $T_0>\tmax$, no topological transition occurs.
Otherwise, define the incoming interface
\begin{equation*}
\Gammabf(T_0^-)
\coloneqq
\lim_{t\uparrow T_0}
\Gammabf(t).
\end{equation*}
Once the surgery constructs the outgoing interface
$\Gammabf(T_k^+)$ and the classical evolution is restarted from
these data, the next event time is
defined inductively by
\begin{equation}\label{event-time-induction}
T_{k+1}
\coloneqq
\inf
\left\{
t\in(T_k,\tmax]:
\reach
\bigl(\Gammabf(t)\bigr)
\leq
\Clyr\eps
\right\}.
\end{equation}
\end{subequations}
If this set is nonempty, set
\begin{equation*}
\Gammabf(T_{k+1}^-)
\coloneqq
\lim_{t\uparrow T_{k+1}}
\Gammabf(t).
\end{equation*}
\end{definition}

Thus, on every open interval $(T^+_k,T^-_{k+1})$,
the interface evolves by the classical Mullins--Sekerka system.
The sharp-interface evolution is suspended only at the
times $T_k$ when the separated-layer condition fails.
At such a time, the loss of separation is
localized by identifying points in the effective transition layer
with more than one nearest point on the interface.

\begin{definition}[Reach-defect set]
\label{def:reach-defect-set}
For a closed interface $\Gammabf$, define the
nearest-point set and medial axis by
\begin{subequations}\label{reach-defect-geometry}
\begin{align}
\Pi_{\Gammabf}(x)
&\coloneqq
\left\{
y\in\Gammabf:
|x-y|=\dist(x,\Gammabf)
\right\}, 
\label{nearest-point-set}\\
\operatorname{Med}(\Gammabf)
&\coloneqq
\left\{
x\in\R^d:
\#\Pi_{\Gammabf}(x)>1
\right\}.
\label{medial-axis}
\end{align}
At the event time $T_k$, define the
\emph{reach-defect set} by
\begin{equation}\label{reach-defect-set}
\Zdfct
\coloneqq
\left\{
x\in\operatorname{Med}
\bigl(\Gammabf(T_k^-)\bigr):
\dist
\bigl(x,\Gammabf(T_k^-)\bigr)
\leq
\Clyr\eps
\right\}.
\end{equation}
\end{subequations}
It contains the points at which the closest-point projection
ceases to be single-valued within the effective transition layer.
\end{definition}

We consider topological events that are
\emph{intermittent} in time and \emph{isolated} in space.  Intermittency
means that only finitely many events occur and that
successive events remain separated on the macroscopic time scale.
Spatial isolation means that each reach-defect set is
contained in a bounded region whose diameter is
$\O(\eps)$.  This models a point-like
coalescence or pinch-off in which the defect-region length
scale is of size $\eps$, i.e., a \emph{microscopic defect}.  
The corresponding $d$-dimensional
volume is $\O(\eps^d)$.  Extended near-contact
regions and simultaneous nonlocal topology changes are excluded.
By Theorem~\ref{thm:CHY}, each classical branch becomes
smooth instantaneously.  Since successive event times remain separated,
we may therefore take every incoming interface to be
a finite union of pairwise disjoint, $C^\infty$-smooth,
embedded hypersurfaces.
These requirements are collected in the following assumption.

\begin{assumption}[Intermittent and isolated transitions]
\label{ass:topology}
There exist an integer $K\in\mathbb{N}_0$ and constants
$\tau_\ast>0$, $C_\ast>0$, and $\cbuf>0$, independent of $\eps$, such
that
\begin{subequations}\label{intermittent-isolated}
\begin{equation}\label{finite-events}
0<T_0<T_1<\cdots<T_K<\tmax,
\qquad
T_{k+1}-T_k\geq\tau_\ast,
\quad 0\leq k<K.
\end{equation}
For each $0\leq k\leq K$, the number of interfaces is finite, $N(T_k^-) < \infty$, and
\begin{equation}\label{incoming-regularity}
X_j(\cdot,T_k^-):\S^{d-1}\longrightarrow\R^d
\text{ is a $C^\infty$ embedding},
\quad
1\leq j\leq N(T_k^-),
\end{equation}
with pairwise disjoint components:
\begin{equation}\label{incoming-separation}
\Gamma_j(T_k^-)\cap\Gamma_\ell(T_k^-)=\emptyset,
\qquad
1\leq j,\ell\leq N(T_k^-),\quad j\neq\ell.
\end{equation}
Moreover, there exists a bounded open defect region $\OmegaCHm$ such
that
\begin{equation}\label{isolated-defect}
\overline{\Zdfct}\subset\OmegaCHm,
\qquad
\diam(\OmegaCHm)\leq C_\ast\eps,
\qquad
\dist\left(\overline{\Zdfct},\p\OmegaCHm\right)
\geq\cbuf\eps.
\end{equation}
\end{subequations}
\end{assumption}

\begin{remark}[Multiple defect regions]
\label{rem:multiple-defect-regions}
\Cref{ass:topology} extends directly to finitely many simultaneous,
spatially separated defects. If
\begin{equation*}
\Zdfct
=
\bigcup_{\ell=1}^{L_k}\Z_{k,\ell}^{\eps},
\qquad
\OmegaCHm
=
\bigcup_{\ell=1}^{L_k}\Omega_{k,\ell}^{\eps},
\qquad
L_k<\infty,
\end{equation*}
where the $\Z_{k,\ell}^{\eps}$ are the connected components of
$\Zdfct$ and the open sets $\Omega_{k,\ell}^{\eps}$ have pairwise
disjoint closures, we require each pair
$(\Z_{k,\ell}^{\eps},\Omega_{k,\ell}^{\eps})$ to satisfy
\eqref{isolated-defect}. If $L_k$ is bounded independently of $\eps$,
then $|\OmegaCHm|=\O(\eps^d)$.
\end{remark}


\section{Localized pseudo-time relaxation}
\label{sec:localized-relaxation}

At each topological event, we suspend the sharp-interface evolution and
resolve the overlapping transition layers in an $\O(\eps)$ neighborhood
of the reach-defect set. This neighborhood is rescaled to a fixed patch,
where a mass-conserving localized Cahn--Hilliard flow is evolved in fast
pseudo-time while the exterior interface remains fixed.

In the following, we suppress the indices $\eps$
and $k$ from the reach-defect set and defect region, writing 
$\Z$ and $\OmegaCH$ for
$\Zdfct$ and $\OmegaCHm$, respectively.
We take $\OmegaCH$ to be a canonical bounded open
domain containing the entire reach-defect set.  More precisely,
we choose a center $x_c\in\mathbb{R}^d$ and a constant
$L>0$, independent of $\eps$, and set
\begin{subequations}\label{canonical-patch}
\begin{equation}
\OmegaCH
=
x_c
+
\eps
(-L,L)^d.
\end{equation}
Thus, $\OmegaCH$ is a square in two dimensions and
a cube in three dimensions.  In accordance with
\eqref{isolated-defect}, it is chosen so that
\begin{equation}\label{canonical-defect-domain}
\overline{\Z}
\subset
\OmegaCH,
\qquad
\dist
\left(
\overline{\Z},
\p\OmegaCH
\right)
\geq
\cbuf\eps,
\qquad
\diam(\OmegaCH)
\leq
C_\ast\eps,
\qquad
|\OmegaCH|
\leq
C_\ast\eps^d.
\end{equation}
\end{subequations}

\begin{definition}[Defect-region cutoff]
\label{def:cutoff}
A \emph{defect-region cutoff} is a smooth compactly supported function
$\eta\in C_c^\infty(\OmegaCH)$ satisfying
\begin{equation}\label{cutoff-properties}
0\leq\eta\leq1
\quad\text{in }\OmegaCH,
\qquad
\eta\equiv1
\quad\text{near }\overline{\Z},
\qquad
\|\p_x^r\eta\|_{L^\infty(\OmegaCH)}
\leq C_r\eps^{-r}, \ r \ge 0, 
\end{equation}
where the constants $C_r$ are independent of $\eps$.
\end{definition}

\subsection{Patch coordinates}

The physical diameter of $\OmegaCH$ is
$\O(\eps)$.  We therefore recenter the patch
at $x_c$ and measure distances in units of $\eps$.
This maps the microscopic physical patch to a fixed
domain of order-one size.

\begin{definition}[Patch coordinates]
\label{def:patch-coords}
The patch coordinate and rescaled defect domain are
\begin{equation}\label{patch-coords}
\xi
\coloneqq
\frac{x-x_c}{\eps},
\qquad
x=x_c+\eps\xi,
\qquad 
\text{and}
\qquad
Q
\coloneqq
\frac{\OmegaCH-x_c}{\eps}
=
(-L,L)^d.
\end{equation}
The rescaled reach-defect set and cutoff are
\begin{equation}\label{rescaled-cutoff}
\mathcal{Y}
\coloneqq
\frac{\mathcal{Z}-x_c}{\eps}
\qquad
\text{and}
\qquad
\etah(\xi)
\coloneqq
\eta(x_c+\eps\xi),
\qquad
\xi\in Q.
\end{equation}
\end{definition}

The chain rule and \eqref{cutoff-properties} imply that
\begin{equation*}
\p_\xi^r\etah(\xi)
=
\eps^r
\p_x^r\eta(x_c+\eps\xi),
\qquad
\|\p_\xi^r\etah\|_{L^\infty(Q)}
\leq
C_r,
\qquad
r\geq0.
\end{equation*}
Moreover, \eqref{canonical-defect-domain} gives
\begin{equation*}
\dist
\left(
\overline{\mathcal{Y}},
\p Q
\right)
\geq
\cbuf.
\end{equation*}
Finally, since $\etah\in C_c^\infty(Q)$, its support is a
compact subset of $Q$ and therefore lies a positive distance from
$\p Q$. Set
\begin{subequations}\label{CH-localized-boundary-collar}
\begin{equation}
\delta_{\etah}
\coloneqq
\frac{1}{2}
\dist
\left(
\operatorname{supp}\etah,
\p Q
\right)
>0,
\end{equation}
and define the corresponding boundary collar by
\begin{equation}
\mathcal{C}_{\etah}
\coloneqq
\left\{
\xi\in Q:
\dist(\xi,\p Q)<\delta_{\etah}
\right\}
\subset
Q\setminus\operatorname{supp}\etah
\subset
\{\etah=0\}.
\end{equation}
\end{subequations}

\subsection{Pseudo-time scaling}

The spatial rescaling determines the corresponding fast
temporal scale.  At the topological event time $T$,
we introduce the inner phase field and chemical potential
\begin{subequations}\label{inner-variables}
\begin{align}
\Phi(\xi,s)
&\coloneqq
\phi_\eps
\left(
x_c+\eps\xi,
T+t_{\mathrm{mix}}s
\right),
\label{inner-phi}
\\
U(\xi,s)
&\coloneqq
\frac{\eps}{\sigmah}
\mu_\eps
\left(
x_c+\eps\xi,
T+t_{\mathrm{mix}}s
\right).
\label{inner-mu}
\end{align}
\end{subequations}
Here, $s$ is the dimensionless pseudo-time,
while $t_{\mathrm{mix}}$ is its physical timescale.
By the chain rule, \eqref{CH-phi} becomes
\begin{align*}
\p_s\Phi
&=
\frac{m_\eps\sigmah t_{\mathrm{mix}}}{\eps^3}
\Delta_\xi U. 
\end{align*}
The two sides balance at order one when
\begin{equation}\label{mixing-scale}
t_{\mathrm{mix}}
=
\frac{\eps^3}{m_\eps\sigmah},
\qquad
s
=
\frac{m_\eps\sigmah}{\eps^3}(t-T).
\end{equation}
In these variables, the inner Cahn--Hilliard system becomes
\begin{subequations}\label{inner-system}
\begin{align}
\p_s\Phi
&=
\Delta_\xi U,
\label{inner-phi-eq}
\\
U
&=
-\Delta_\xi\Phi+W'(\Phi).
\label{inner-mu-eq}
\end{align}
\end{subequations}

Thus, an $\O(1)$ pseudo-time interval corresponds to
an $\O(\eps^3)$ physical-time interval.  An
order-one macroscale velocity displaces the interface by only
$\O(\eps^3)$ during this interval, which is negligible
relative to the $\O(\eps)$ defect-region length scale.
The macroscale sharp-interface evolution may therefore be frozen
at $t=T$ during the localized pseudo-time evolution.

The spatial rescaling also transforms the free energy.
The contribution from the physical defect region is
\begin{equation}\label{defect-eng}
\Eng_{\eps,\OmegaCH}[\phi_\eps]
\coloneqq
\sigmah
\int_{\OmegaCH}
\left(
\frac{\eps}{2}|\nabla_x\phi_\eps|^2
+\frac{1}{\eps}W(\phi_\eps)
\right)
\dd{x} 
=
\sigmah\eps^{d-1}\Feng[\Phi], 
\end{equation}
where $\Feng[\Phi]$ is the rescaled free energy associated with 
\eqref{inner-system}, 
\begin{equation}\label{rescaled-defect-eng}
\Feng[\Phi]
\coloneqq
\int_Q
\left(
\frac{1}{2}|\nabla_\xi\Phi|^2
+W(\Phi)
\right)
\dd{\xi}.
\end{equation}

\subsection{Localized dissipation form}

The system \eqref{inner-system} has
the same gradient-flow structure as \eqref{CH}.
More precisely, it is the $H^{-1}(Q)$-gradient flow
of the rescaled free energy \eqref{rescaled-defect-eng}.  
Here, the $H^{-1}(Q)$ inner product is defined as in
\eqref{H1-inner-product}, with the whole-space operator $(-\Delta)^{-1}$ replaced by the inverse
Neumann Laplacian on zero-mean functions in $Q$.  
The dissipation law for \eqref{inner-system} is
\begin{equation}\label{inner-eng-dissipation}
\frac{\dd{\,}}{\dd s}
\Feng[\Phi]
=
-\int_Q
|\nabla_\xi U|^2
\dd{\xi}.
\end{equation}
Thus, the classical local Cahn--Hilliard evolution \eqref{inner-system} 
dissipates energy wherever $\nabla_\xi U\neq0$ in $Q$. 
The interface outside $Q$ remains fixed; if dissipation 
extends to $\partial Q$, the zero level set will move
and therefore fail to join smoothly to the
unchanged exterior interface.
To restrict the dissipative effects to the defect region,
we employ the cutoff \eqref{rescaled-cutoff} and 
introduce the following symmetric bilinear form.

\begin{definition}[Localized dissipation form]
\label{def:localized-form}
For smooth functions $f,g:Q\to\mathbb{R}$, define
\begin{subequations}\label{localized-dissipation-form}
\begin{equation}\label{localized-bilinear-form}
\ain_{\etah}(f,g)
\coloneqq
\int_Q
\nabla_\xi(\etah f)
\cdot
\nabla_\xi(\etah g)
\dd{\xi}.
\end{equation}
The associated localized dissipation seminorm is
\begin{equation}\label{localized-dissipation-seminorm}
|f|_{\etah}
\coloneqq
\ain_{\etah}(f,f)^{1/2}
=
\left\|
\nabla_\xi(\etah f)
\right\|_{L^2(Q)}.
\end{equation}
\end{subequations}
\end{definition}

The form \eqref{localized-bilinear-form} is symmetric and positive
semidefinite, while \eqref{localized-dissipation-seminorm} defines only
a seminorm. The resulting degeneracy will freeze the evolution where
$\etah=0$. By \eqref{cutoff-properties},
$\etah=1$ on the defect core, recovering the usual
Cahn--Hilliard dissipation.

\subsection{Mass-conserving localized evolution}

Next, we use the localized dissipation form \eqref{localized-dissipation-form} to 
construct a mass-conserving gradient flow whose 
dissipation is confined to the reach-defect region.
Let $\Phi_{\mathrm{in}}\coloneqq\Phi(\cdot,0)$. 
By \eqref{sdf-phi},
\begin{equation*}
\Phi_{\mathrm{in}}(\xi)
=
q\bigl(D_{\mathrm{in}}(\xi)\bigr),
\qquad
D_{\mathrm{in}}(\xi)
\coloneqq
\frac{1}{\eps}
d_{\Gammabf(T^-)}(x_c+\eps\xi).
\end{equation*}
It follows that
\begin{equation*}
|\Phi_{\mathrm{in}}(\xi)-\Phi_{\mathrm{in}}(\zeta)|
\leq
\|q'\|_{L^\infty}
|\xi-\zeta|, 
\end{equation*}
so that $\Phi_{\mathrm{in}}\in W^{1,\infty}(Q)\subset H^1(Q)$.
On the rescaled reach-defect set $\mathcal{Y}$, however,
the closest-point projection is not unique, so
$D_{\mathrm{in}}$ and $\Phi_{\mathrm{in}}$ are generally not
continuously differentiable. 
The cutoff then restricts the set of admissible configurations to
\begin{equation}\label{admissible-class}
\mathcal{X}
\coloneqq
\left\{
\Psi\in H^1(Q):
\Psi=\Phi_{\mathrm{in}}
\text{ a.e. on }\{\etah=0\}
\right\}.
\end{equation}
The pseudo-time evolution is associated with the constrained
minimization problem
\begin{equation}\label{local-minimization}
\min_{\Psi\in\mathcal{X}}
\Feng[\Psi]
\qquad
\text{subject to}
\qquad
\int_Q\Psi\dd{\xi}
=
\int_Q\Phi_{\mathrm{in}}\dd{\xi}.
\end{equation}
The corresponding Lagrangian is
\begin{equation}\label{lagrangian-constrained}
\mathcal{L}[\Psi,\lambda]
\coloneqq
\Feng[\Psi]
+
\lambda
\left(
\int_Q\Psi\dd{\xi}
-\int_Q\Phi_{\mathrm{in}}\dd{\xi}
\right),
\end{equation}
where $\lambda\in\mathbb{R}$ is the Lagrange multiplier
for the phase-mass constraint.  Since the variational
derivative of \eqref{lagrangian-constrained} is
$U+\lambda$, the localized constrained gradient flow is
defined weakly by
\begin{equation}\label{local-weak-flow}
\int_Q
\p_s\Phi\,\psi
\dd{\xi}
=
-\ain_{\etah}(U+\lambda,\psi),
\qquad
\psi\in H^1(Q).
\end{equation}
Here, $\lambda=\lambda(s)$ is chosen so that the
flow remains on the constraint set in
\eqref{local-minimization}.  Differentiating the constraint
gives
\begin{equation*}
\frac{\dd{\,}}{\dd s}
\int_Q\Phi\dd{\xi}
=0
\implies 
\ain_{\etah}(U+\lambda,1)
=0.
\end{equation*}
Since the cutoff is nonconstant,
\begin{equation*}
\ain_{\etah}(1,1)
=
\int_Q|\nabla_\xi\etah|^2\dd{\xi}
>0,
\end{equation*}
so the multiplier is uniquely determined as
\begin{equation}
\lambda(s) = - \frac{\ain_{\etah}(U,1)}{\ain_{\etah}(1,1)}. 
\end{equation}

\subsection{Localized pseudo-time system}

We now write the localized weak flow 
\eqref{local-weak-flow} in strong form.  Define the
localized chemical potential
\begin{equation}\label{localized-mu}
Z
\coloneqq
(U+\lambda)\etah.
\end{equation}
Since $\etah\in C_c^\infty(Q)$, the product
$\etah\psi$ belongs to $H_0^1(Q)$.  Assuming sufficient
regularity, integration by parts in
\eqref{local-weak-flow} gives
\begin{align*}
-\ain_{\etah}(U+\lambda,\psi)
=
-\int_Q
\nabla_\xi Z
\cdot
\nabla_\xi(\etah\psi)
\dd{\xi}
=
\int_Q
\etah\Delta_\xi Z\,\psi
\dd{\xi}.
\end{align*}
Consequently, the localized pseudo-time evolution is
\begin{subequations}\label{localized-eqs}
\begin{align}
\p_s\Phi
&=
\etah\Delta_\xi Z,
&&
\text{in }Q\times(0,\smax),
\label{localized-eqs-phi}
\\
Z
&=
(U+\lambda)\etah,
&&
\text{in }Q\times(0,\smax),
\label{localized-eqs-z}
\\
U
&=
-\Delta_\xi\Phi+W'(\Phi),
&&
\text{in }Q\times(0,\smax),
\label{localized-eqs-mu}
\\
\lambda(s)
&=
-\frac{
\int_Q
\nabla_\xi(\etah U)
\cdot
\nabla_\xi\etah
\dd{\xi}
}{
\int_Q
|\nabla_\xi\etah|^2
\dd{\xi}
},
&&
0<s<\smax,
\label{localized-eqs-lambda}
\\
\Phi(\xi,0)
&=
\Phi_{\mathrm{in}}(\xi),
&&
\text{in }Q,
\label{localized-eqs-init}
\\
\Phi(\xi,s)
&=
\Phi_{\mathrm{in}}(\xi),
\qquad
Z(\xi,s)=0,
&&
\text{in }\mathcal{C}_{\etah}\times(0,\smax).
\label{localized-eqs-bcs}
\end{align}
\end{subequations}
Here, $\smax$ is the stopping pseudo-time, which will be defined below.
The collar identities
\eqref{localized-eqs-bcs} are not
independent boundary conditions.  Since $\etah=0$ on
$\mathcal{C}_{\etah}$,
\eqref{localized-eqs-phi} gives
$\p_s\Phi=0$ there, and the initial condition then gives
$\Phi=\Phi_{\mathrm{in}}$; similarly,
\eqref{localized-eqs-z} gives $Z=0$.
Thus the localized evolution itself freezes both fields throughout
an open neighborhood of $\p Q$.
For numerical purposes, it is convenient to replace 
\eqref{localized-eqs-bcs} by 
the weaker implied Neumann conditions, 
\begin{equation}\label{localized-eqs-bcs-neumann}
\p_\nu
\bigl(
\Phi-\Phi_{\mathrm{in}}
\bigr)
=
0,
\qquad
\p_\nu Z
=
0,
\qquad
\text{on }\p Q\times(0,\smax),
\end{equation}
where $\nu$ is the outward unit normal to $\p Q$. 

Taking $\psi=U+\lambda$ in
\eqref{local-weak-flow} and using the
phase-mass constraint gives the localized energy dissipation law
\begin{equation}\label{localized-eqs-dissipation}
\frac{\dd{\,}}{\dd s}
\Feng[\Phi(s)]
=
-\ain_{\etah}(U+\lambda,U+\lambda)
=
-\int_Q|\nabla_\xi Z|^2\dd{\xi}
\eqqcolon
-\mathcal{D}_{\etah}(s).
\end{equation}

\subsection{Termination of the localized relaxation}

The localized system \eqref{localized-eqs} 
is evolved only while it produces an
appreciable decrease of the free energy.  
For nontrivial interface surgery, the localized free energy is positive, 
$\Feng[\Phi(s)]>0$. We define the relative energy-dissipation rate by
\begin{equation}\label{dissipation-rate}
\mathcal{R}_{\etah}(s)
\coloneqq
\frac{
\mathcal{D}_{\etah}(s)
}{
\Feng[\Phi(s)]
}
=
-\frac{\dd{\,}}{\dd s}
\log\Feng[\Phi(s)],
\end{equation}
where $\mathcal{D}_{\etah}$ is defined in \eqref{localized-eqs-dissipation}. 
Thus $\mathcal{R}_{\etah}$ measures the instantaneous fractional
rate at which the free energy decreases.

\begin{definition}[Terminal pseudo-time]
\label{def:stopping-time}
Fix a dissipation tolerance $\delta_s>0$.  The
localized pseudo-time evolution \eqref{localized-eqs} is terminated at
the terminal pseudo-time
\begin{equation}\label{stopping-time} 
\smax
\coloneqq
\inf
\left\{
s>0:
\mathcal{R}_{\etah}(s)
\leq
\delta_s
\right\}. 
\end{equation}
\end{definition}

Let $\Feng_{\min}$ denote the mass-constrained
infimum of the rescaled free energy, 
\begin{equation}\label{min-eng}
\Feng_{\min}
\coloneqq
\inf
\left\{
\Feng[\Psi]:
\Psi\in\mathcal{X},
\quad
\int_Q\Psi\dd{\xi}
=
\int_Q\Phi_{\mathrm{in}}\dd{\xi}
\right\}.
\end{equation}

\begin{lemma}[Bound on the terminal pseudo-time]
\label{lem:stopping-time-bound}
Suppose that there are constants $c_F,C_F>0$, independent of
$\eps$, such that
\begin{equation}\label{eng-bounds}
0<c_F
\leq
\Feng_{\min}
\leq
\Feng[\Phi_{\mathrm{in}}]
\leq
C_F.
\end{equation}
Then the terminal pseudo-time $\smax$ defined by \eqref{stopping-time} is bounded above by
\begin{equation}\label{stopping-time-bound}
\smax
\leq
\frac{1}{\delta_s}
\log
\left(
\frac{C_F}{c_F}
\right).
\end{equation}
Consequently, $\smax=\mathcal{O}(1)$ as $\eps \to 0$.
\end{lemma}

\begin{proof}
For $0<s<\smax$, \eqref{stopping-time} gives
$\mathcal{R}_{\etah}(s)>\delta_s$.  Hence,
by \eqref{dissipation-rate},
\begin{equation*}
\Feng[\Phi(s)]
\leq
\Feng[\Phi_{\mathrm{in}}]
e^{-\delta_s s}.
\end{equation*}
Combining this inequality with
\eqref{eng-bounds} gives
\eqref{stopping-time-bound}.
\end{proof}


\section{Interface surgery and sharp-interface restart}
\label{sec:surgery}

At each topological event, the localized system
\eqref{localized-eqs} is evolved to the terminal pseudo-time
$s=\smax$ defined in \eqref{stopping-time}, and the terminal phase field
$\Phi(\cdot,\smax)$ is used to construct an outgoing interface from
which the sharp-interface evolution is restarted. A rigorous
justification requires two ingredients: (1) the localized system must be
globally well posed and relax toward a sufficiently regular, stable
equilibrium, and (2)  the zero level set
$\{\Phi(\cdot,\smax)=0\}$ must define a smooth outgoing interface that
matches the incoming interface outside the defect region and again
satisfies \eqref{separated-layers}. Complete proofs of these properties
lie beyond the scope of this work; below, we outline the basic
analytical framework and verify the geometric reconstruction for an
idealized coalescence configuration.

\subsection{Well-posedness and asymptotic relaxation}

The standard Galerkin and energy-estimate arguments for the
Cahn--Hilliard system \eqref{CH} (see, e.g., Miranville~\cite{Miranville2019}) 
can be adapted to establish
well-posedness of the localized system \eqref{localized-eqs} in a
natural weak-solution class. The principal additional difficulty is the
cutoff-induced degeneracy: the leading fourth-order coefficient is
$\etah^2$ and vanishes where $\etah=0$, so the equation is not uniformly
parabolic throughout $Q$. Nevertheless, $\etah$ is bounded away from
zero on every compact subset of $\{\etah>0\}$ and is identically one
near the rescaled reach-defect set $\mathcal{Y}$. Standard interior
regularity therefore applies in the region where the topological
transition is resolved.

The dissipation identity \eqref{localized-eqs-dissipation} makes
$\Feng$ a Lyapunov functional for the localized evolution. For the
classical Cahn--Hilliard equation, compactness and
{\L}ojasiewicz--Simon arguments yield convergence to a single equilibrium
\cite{RyHo1999}. The same general framework provides the natural route
to describing the long-time behavior of \eqref{localized-eqs}, although
the cutoff-induced degeneracy must be accommodated in the corresponding
functional setting. Suppose that the localized evolution converges to an isolated, stable
constrained minimizer $\Phi_\ast$. If the stopping criterion places
$\Phi(\cdot,\smax)$ sufficiently close to $\Phi_\ast$ in a topology
strong enough to control level-set geometry, then the smoothness,
matching, and reach properties of $\{\Phi_\ast=0\}$ transfer to the
terminal zero level set $\{\Phi(\cdot,\smax)=0\}$.  
Thus, proving that the equilibrium zero-level set satisfies the regularity requirement
\eqref{CHY-initial-regularity} and restores the separated-layer condition
in \Cref{ass:separated-layers} would justify the sharp-interface restart, 
completing the logical chain underlying the sharp--diffuse continuation methodology.

\subsection{Smooth interface surgery for idealized coalescence}

We examine these two requirements in an idealized symmetric model of
interface coalescence. The symmetry permits an explicit computation of the post-coalescence
sharp-interface equilibrium. We show that this interface satisfies the
reach condition \eqref{separated-layers} and matches the prescribed
exterior interface with $C^4$ regularity at the attachment points.

\subsubsection{Incoming idealized interface}

We specialize to $d=2$ and take
\begin{equation}\label{ideal-domain}
Q=(-L,L)^2.
\end{equation}
For this idealized calculation, we suppress the cutoff transition and
set $\etah\equiv1$. The localized evolution then becomes the unweighted
inner Cahn--Hilliard system
\begin{subequations}\label{ideal-system}
\begin{align}
\p_s\Phi
&=
\Delta_\xi U,
\\
U
&=
-\Delta_\xi\Phi+W'(\Phi),
\end{align}
with homogeneous Neumann conditions
\begin{equation}\label{ideal-bcs}
\p_\nu\Phi=0,
\qquad
\p_\nu U=0
\qquad
\text{on }\p Q.
\end{equation}
\end{subequations}
Here the phase mass is conserved automatically, so no additional
multiplier is required. 

The incoming interface consists of the two graphs
\begin{equation}\label{ideal-incoming}
\Gamma_\pm^{\mathrm{in}}
\coloneqq
\left\{
(\xi_1,\pm h(\xi_1)):
-L\leq\xi_1\leq L
\right\}
\subset\overline Q,
\end{equation}
where $h\in C^4([-L,L])$ is positive and
\begin{equation}\label{ideal-attachment}
h(\pm L)=H.
\end{equation}
Thus the four attachment points are $(\pm L,\pm H)$.

\begin{lemma}[Neumann conditions at the attachment points]
\label{lem:ideal-orthogonal-contact}
Let $\Gamma$ be a smooth extension of
$\Gamma_+^{\mathrm{in}}\cup\Gamma_-^{\mathrm{in}}$ through the four
attachment points, and suppose that its signed-distance function
$d_\Gamma$ is smooth there. Define
\begin{equation*}
\Phi_{\mathrm{in}}
=
q(d_\Gamma),
\qquad
U_{\mathrm{in}}
=
-\Delta_\xi\Phi_{\mathrm{in}}
+
W'(\Phi_{\mathrm{in}}).
\end{equation*}
If $(\Phi_{\mathrm{in}},U_{\mathrm{in}})$ satisfies
\eqref{ideal-bcs}, then the incoming branches meet the vertical sides
of $Q$ orthogonally and
\begin{equation}\label{ideal-orthogonal-contact}
h'(\pm L)=0,
\qquad
h'''(\pm L)=0.
\end{equation}
\end{lemma}

\begin{proof}
At an attachment point,
\begin{equation*}
\p_\nu\Phi_{\mathrm{in}}
=
q'(0)\nabla_\xi d_\Gamma\cdot\nu
=
q'(0)n\cdot\nu.
\end{equation*}
Since $q'(0)\neq0$, the first condition in \eqref{ideal-bcs} gives
$n\cdot\nu=0$. Thus the interface meets the vertical side
orthogonally, so that  $h'(\pm L)=0$.
Wherever $d_\Gamma$ is smooth, $|\nabla_\xi d_\Gamma|=1$, and hence
\begin{align*}
U_{\mathrm{in}}
&=
-q''(d_\Gamma)
-q'(d_\Gamma)\Delta_\xi d_\Gamma
+
W'\bigl(q(d_\Gamma)\bigr)
\\
&=
-q'(d_\Gamma)\Delta_\xi d_\Gamma,
\end{align*}
where the profile equation \eqref{profile-eq} was used in the second equality. 
At an attachment point, $\p_\nu d_\Gamma=n\cdot\nu=0$, so
\begin{equation*}
\p_\nu U_{\mathrm{in}}
=
-q'(0)\p_\nu\Delta_\xi d_\Gamma.
\end{equation*}
The second condition in \eqref{ideal-bcs} therefore gives
$\p_\nu\Delta_\xi d_\Gamma=0$. Since $\nu$ is tangent to the interface
at orthogonal contact, this is the vanishing of the tangential
derivative of curvature. For a graph satisfying $h'=0$, that derivative
is $h'''$.
\end{proof}

We impose the remaining geometric conditions as follows.

\begin{assumption}[Idealized coalescence]
\label{ass:ideal-coalescence}
Let $h$ satisfy the attachment conditions
\eqref{ideal-attachment} and \eqref{ideal-orthogonal-contact}, and
suppose that
\begin{equation}\label{ideal-reach-compatible-scales}
0 < h(0) < \Clyr<H<L-\Clyr.
\end{equation}
The function $h$ is assumed to be even and to satisfy
\begin{equation}\label{ideal-centre-conditions}
h'(0)=0,
\qquad
h''(0)>0,
\end{equation}
together with the attachment conditions
\begin{equation}\label{ideal-boundary-conditions}
h''(\pm L)
=
-\frac{1}{H},
\qquad
h^{(4)}(\pm L)
=
-\frac{3}{H^3}.
\end{equation}
Finally, we impose the area constraint
\begin{equation}\label{ideal-mass-constraint}
\int_0^L h(\xi_1)\dd{\xi_1}
=
\frac{\pi}{4}H^2.
\end{equation}
\end{assumption}

The two incoming branches therefore have a nondegenerate closest
approach at $\xi_1=0$ and meet the vertical sides of $Q$ orthogonally
at $(\pm L,\pm H)$. Since their separation at the center is
$2h(0)<2\Clyr$, their effective transition layers overlap there. The
constraint \eqref{ideal-mass-constraint} equates the area between the
incoming graphs with the total area enclosed by the two outgoing
semicircles constructed below.

\begin{figure}[tbhp]
\centering
\includegraphics[width=0.4\textwidth]{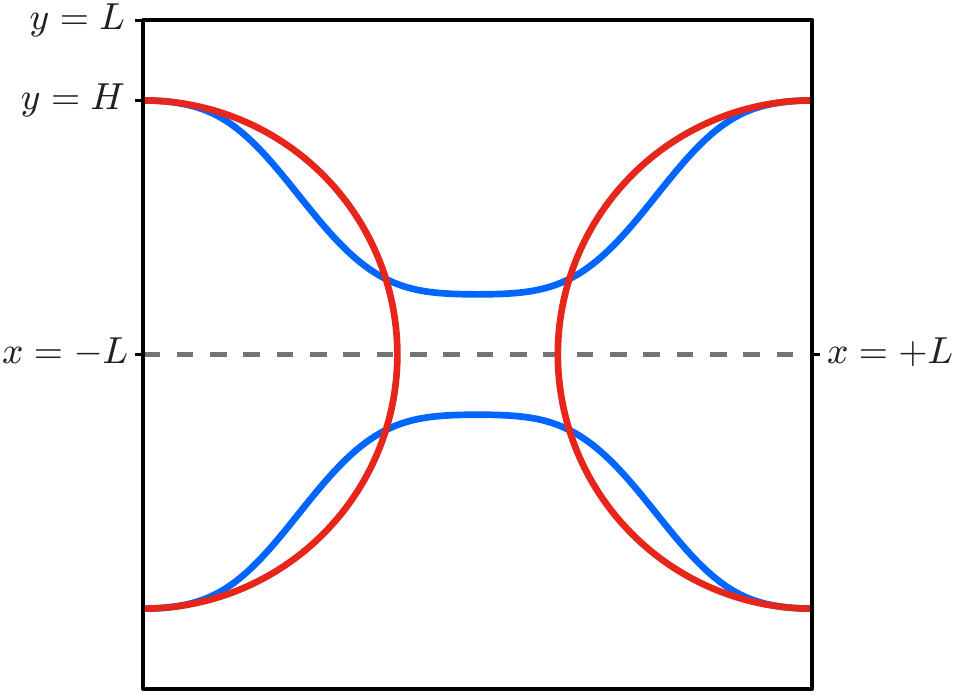}
\caption{
\textbf{Idealized coalescence.}
The blue curves are the incoming interface branches
$\xi_2=\pm h(\xi_1)$ generated by the polynomial profile
\eqref{ideal-polynomial-profile} with the parameter choice
\eqref{ideal-example-parameters}. The red curves are the reconstructed
outgoing equilibrium semicircles \eqref{ideal-outgoing}.
}
\label{fig:ideal}
\end{figure}

\begin{example}[Explicit polynomial height function]
\label{ex:ideal-polynomial-profile}
\begin{subequations}\label{ideal-polynomial-profile}
For prescribed values of $L$, $H$, and $h(0)$ satisfying
\eqref{ideal-reach-compatible-scales}, 
we construct an explicit admissible height function.
Introduce the dimensionless aspect ratio and central height, 
\begin{equation}\label{ideal-polynomial-variables}
\Lambda
\coloneqq
\frac{L}{H},
\qquad
\alpha
\coloneqq
\frac{h(0)}{H}. 
\end{equation}
Set
$r\coloneqq1-{\xi_1^2}/{L^2}$ and define the height function by
\begin{equation}\label{ideal-polynomial-h}
h(\xi_1)
=
H\mathcal{P}(r),
\end{equation}
where
\begin{equation}\label{ideal-polynomial-P}
\mathcal{P}(r)
={}
1
-\frac{\Lambda^2}{8}r^2
-\frac{\Lambda^2}{16}r^3
-\frac{\Lambda^4+5\Lambda^2}{128}r^4
+Ar^5
+Br^5(1-r).
\end{equation}
The terms through order $r^4$ enforce the attachment conditions. The
coefficient $A$ enforces the prescribed central height, 
\begin{equation}\label{ideal-polynomial-A}
A
=
\alpha-1
+\frac{\Lambda^4+29\Lambda^2}{128}, 
\end{equation}
and the coefficient $B$ is determined by the area constraint
\eqref{ideal-mass-constraint}, 
\begin{equation}\label{ideal-polynomial-B}
B
=
\frac{9009}{256}
\left(
\frac{\pi}{4\Lambda}
-1
+\frac{\Lambda^2}{9}
+\frac{\Lambda^4}{315}
-\frac{256}{693}A
\right).
\end{equation}
\end{subequations}

The resulting function $h$ is an even polynomial in $\xi_1$ of degree
at most twelve. It satisfies the attachment conditions
\eqref{ideal-attachment}, \eqref{ideal-orthogonal-contact}, and
\eqref{ideal-boundary-conditions}, the area constraint
\eqref{ideal-mass-constraint}, and the central conditions
$h(0)=H\alpha$ and $h'(0)=0$.
The sufficient conditions
\begin{equation}\label{ideal-polynomial-admissibility}
1<\Lambda<\frac{5}{2},
\qquad
\frac{\Lambda^4+19\Lambda^2}{32}
-5A+B
>0
\end{equation}
ensure that $h$ is positive on $[-L,L]$ and that $h''(0)>0$.
Consequently, under \eqref{ideal-polynomial-admissibility}, the profile
satisfies all the requirements of \Cref{ass:ideal-coalescence}. The
algebraic verification is given in \Cref{app:polynomial-profile}.

In particular, the parameter choice
\begin{equation}\label{ideal-example-parameters}
\Clyr=2,
\qquad
L=10,
\qquad
H=7.5,
\qquad
h(0)=1.5
\end{equation}
satisfies the scale condition
\eqref{ideal-reach-compatible-scales} and the admissibility conditions
\eqref{ideal-polynomial-admissibility}. The corresponding incoming and
outgoing interfaces are shown in \Cref{fig:ideal}.
\end{example}

\subsubsection{Outgoing equilibrium interface}

Each component of the outgoing equilibrium interface $\Gamma_\pm^{\mathrm{out}}$
has constant curvature and is therefore circular. Symmetry about the
horizontal axis, attachment at $(\pm L,\pm H)$, and orthogonal contact
with the vertical sides of $Q$ place the centers at $(\pm L,0)$ and fix
the radius to be $H$. Thus the outgoing interface consists of the two
inward-facing semicircles
\begin{equation}\label{ideal-outgoing}
\Gamma_\pm^{\mathrm{out}}
\coloneqq
\left\{
(\xi_1,\xi_2)\in\overline{Q}:
(\xi_1\mp L)^2+\xi_2^2=H^2
\right\},
\end{equation}
shown in red in \Cref{fig:ideal}. 
Each semicircle has radius $H$, and the distance between the
two components is $2(L-H)$. Consequently,
\begin{equation}\label{ideal-outgoing-reach}
\reach\left(
\Gamma_-^{\mathrm{out}}
\cup
\Gamma_+^{\mathrm{out}}
\right)
=
\min\{H,L-H\}
>
\Clyr,
\end{equation}
where the final inequality follows from
\eqref{ideal-reach-compatible-scales}. Under the physical rescaling
$x=x_c+\eps\xi$, the reach is multiplied by $\eps$, producing the
required bound \eqref{separated-layers}. 
The matching conditions
\eqref{ideal-attachment},
\eqref{ideal-orthogonal-contact}, and
\eqref{ideal-boundary-conditions}
then ensure that the outgoing arcs join the prescribed exterior
interface with $C^4$ regularity.

The reconstructed global outgoing interface is therefore a compact,
embedded $C^4$ hypersurface satisfying the separated-layer condition.
In particular, it belongs to $C^{3+r}$ for every $r\in(0,1)$ and hence
provides admissible initial data for \Cref{thm:CHY}; the sharp-interface
evolution can then be restarted from this interface.


\section{Some details of the numerical implementation}
\label{sec:implementation}

We now describe our simple two-dimensional implementation of the
event-driven sharp--diffuse continuation procedure.\footnote{%
The code was written in MATLAB, and all numerical experiments
were performed on a MacBook Pro equipped with an Apple M1 Pro processor
and 32\,GB of memory.} Between topological
events, the Mullins--Sekerka evolution is computed using the
boundary-integral method of Zhu--Chen--Hou~\cite{ZhChHo1996}; during each localized
relaxation, the Cahn--Hilliard system is discretized using an SAV scheme
adapted from Shen--Xu--Yang~\cite{ShXuYa2018}.

\subsection{Overview of the hybrid algorithm}

The computation alternates between sharp-interface evolution on the
full domain and localized diffuse relaxation near detected topological
events. The complete event-driven procedure is summarized in
\Cref{alg:hybrid}.
\begin{algorithm}[tbhp]
\caption{Event-driven hybrid sharp--diffuse algorithm}
\label{alg:hybrid}

\begin{enumerate}[
label=\textbf{Step \arabic*.},
leftmargin=*,
topsep=0.5em,
itemsep=0.5em
]

\item
\emph{Sharp-interface evolution.}
Advance the interface using the boundary-integral formulation
\eqref{BI-eqs} and \eqref{BI-interface} while the discrete reach
criterion remains satisfied.

\item
\emph{Event detection and localization.}
When the reach-defect detector described in
\Cref{alg:detector} flags an impending
topological event, record the incoming interface
$\Gammabf(T^-)$ and construct the defect patch, cutoff, and rescaled
coordinates according to
\eqref{canonical-patch},
\Cref{def:cutoff}, and
\Cref{def:patch-coords}.

\item
\emph{Localized relaxation.}
Freeze the sharp-interface evolution and initialize
$\Phi_{\mathrm{in}}$ from the signed-distance profile
\eqref{sdf-phi} in the patch coordinates
\eqref{patch-coords}. Discretize the localized system
\eqref{localized-eqs} using a finite-volume SAV scheme and evolve
until the stopping criterion \eqref{stopping-time} is reached.

\item
\emph{Interface reconstruction.}
Extract the terminal zero level set
$\{\Phi(\cdot,\smax)=0\}$ using marching squares, join it to the unchanged exterior
interface, and reparametrize the resulting closed curves for the
boundary-integral solver.

\item
\emph{Sharp-interface restart.}
Advance the physical time according to \eqref{mixing-scale}, verify
the outgoing separated-layer condition, and restart the
sharp-interface evolution from $\Gammabf(T^+)$. Repeat the procedure
until $t=\tmax$.

\end{enumerate}
\end{algorithm}

\subsection{Boundary-integral discretization}

For the sharp-interface evolution, we use the $\theta$--$L$
boundary-integral method developed in \cite{HoLoSh1994,ZhChHo1996}.
Each interface component is represented by
\begin{subequations}\label{theta-L-representation}
\begin{align}
\p_\alpha X_j(\alpha,t)
&=
\frac{L_j(t)}{2\pi}
\begin{pmatrix}
\cos\theta_j(\alpha,t)
\\
\sin\theta_j(\alpha,t)
\end{pmatrix},
\label{theta-L-tangent}
\\
\theta_j(\alpha,t)
&=
\alpha+\psi_j(\alpha,t),
\label{theta-L-angle}
\end{align}
where $\alpha\in[0,2\pi]$, each component is oriented
counterclockwise, $L_j(t)$ is its length, and $\psi_j$ is $2\pi$-periodic.
In particular, since $|\p_\alpha X_j(\alpha,t)| = {L_j(t)} / {2\pi}$,
it follows that $\alpha$ is an equal-arclength parameter.
The variables $\theta_j$ and $L_j$ determine the shape of
$\Gamma_j(t)$ but not its position. We therefore also evolve the reference
point $a_j(t) \coloneqq X_j(0,t)$ and reconstruct the interface from
\begin{equation}\label{interface-reconstruction}
X_j(\alpha,t)
=
a_j(t)
+
\frac{L_j(t)}{2\pi}
\int_0^\alpha
\begin{pmatrix}
\cos\theta_j(\beta,t)
\\
\sin\theta_j(\beta,t)
\end{pmatrix}
\dd{\beta}.
\end{equation}
\end{subequations}
The reference point $a_j(t)$ is advanced using
\eqref{BI-interface}, with the tangential velocity $V_j^\tau$ 
chosen to preserve the equal-arclength parametrization
\eqref{theta-L-representation}. 
We refer the reader to \cite{HoLoSh1994,ZhChHo1996} for further details. 

We use a Fourier spectral discretization in space together with the
small-scale decomposition described in \cite{ZhChHo1996}. Under this
decomposition, the stiff leading-order term in the tangent-angle
equation is diagonal in Fourier space and is treated implicitly, while
the remaining terms are treated explicitly using a first-order IMEX
scheme. We replace the second-order time discretization and
preconditioned conjugate-gradient iteration of \cite{ZhChHo1996} with
this first-order scheme and a direct solve using MATLAB's backslash operator.

\subsection{Event detection and defect-patch construction}

In our numerical implementation, we have replaced direct evaluation of the reach
condition \eqref{separated-layers} with a computable
geometric proxy adapted from the \emph{manifold-death} algorithm of
Chirco et al.~\cite{ChMaPoZa2022}. The proxy detects the impending
overlap of diffuse transition layers at the prescribed microscopic
length scale and localizes the corresponding defect region. The
event-detection procedure is summarized in \Cref{alg:detector}.

\begin{algorithm}[tbhp]
\caption{Approximate reach-defect detection}
\label{alg:detector}

\begin{enumerate}[
label=\textbf{Step \arabic*.},
leftmargin=*,
topsep=0.5em,
itemsep=0.75em
]

\item
\emph{Detection grid and phase indicator.}
Construct the piecewise-linear interface $\Gammabf_h(t)$ and a uniform
Cartesian grid of spacing $h$ in the tubular neighborhood
$\mathcal{N}_{R_{\mathsf{tube}}}(t)$ defined by \eqref{tube}, where 
\begin{equation}\label{detector-scales}
h
=
\frac{2\Clyr\eps}{N_{\mathsf{layer}}},
\qquad
R
=
2\Clyr\eps,
\qquad
R_{\mathsf{tube}}
=
\Clyr\eps+R+\sqrt{2}h.
\end{equation}
Evaluate the signed distance and equilibrium phase field using
\eqref{sdf-def} and \eqref{sdf-phi}, and set
$c_h=\operatorname{sgn}(\phi_h)$. We use
$N_{\mathsf{layer}}=8$, $\Clyr=2$, and $h=\eps/2$.

\item
\emph{Moment signature.}
For each grid point $x_i$ satisfying
$|d_h(x_i,t)|\leq\Clyr\eps$, define the normalized
moment tensor by:
\begin{subequations}\label{detector-tensor}
\begin{align}
\mathsf{T}_{R,h}(x_i)
&\coloneqq
\frac{h^2}{Z_{R,h}}
\sum_{x_\ell\in\mathcal{A}_{R,h}(x_i)}
(x_\ell-x_i)\otimes(x_\ell-x_i)c_h(x_\ell,t),
\\
\mathcal{A}_{R,h}(x_i)
&\coloneqq
\left\{
x_\ell:
\bigl||x_\ell-x_i|-R\bigr|
\leq\tfrac{h}{2}
\right\},
\\
Z_{R,h}
&\coloneqq
\tfrac{\pi}{4}
\left[
\left(R+\tfrac{h}{2}\right)^4
-
\left(R-\tfrac{h}{2}\right)^4
\right].
\end{align}
\end{subequations}
Let $\lambda_{1,h}(x_i)\geq\lambda_{2,h}(x_i)$ be the eigenvalues of
$\mathsf{T}_{R,h}(x_i)$. Classify $x_i$ as \emph{thin} if
\begin{equation}\label{thin-node-criterion}
\lambda_{1,h}(x_i)>\delta_{\mathsf{sig}},
\qquad
\lambda_{2,h}(x_i)<-\delta_{\mathsf{sig}},
\qquad
\delta_{\mathsf{sig}}=10^{-1}.
\end{equation}

\item
\emph{Event time and defect localization.}
The first time at which the set of thin points is
nonempty defines the numerical topological event time. Its connected components
provide discrete approximations of the reach-defect sets. Enclose each
component in a buffered square satisfying
\eqref{canonical-patch} and
\eqref{canonical-defect-domain}; these squares define the defect
regions $\OmegaCH$ used for the localized relaxation.

\end{enumerate}
\end{algorithm}

\subsection{SAV scheme for local relaxation}

The localized relaxation \eqref{localized-eqs} is discretized in space
by a finite-volume method on the uniform patch grid, with the Neumann
conditions \eqref{localized-eqs-bcs-neumann}. The cutoff is applied
symmetrically in the discrete dissipation operator, thereby freezing
the boundary collar, while the discrete counterpart of
\eqref{localized-eqs-lambda} enforces exact conservation of the phase
mass. Pseudo-time integration uses the second-order SAV--BDF2 scheme
described in \cite{ShXuYa2018}. Each step requires only linear solves
with time-independent matrices and dissipates a modified discrete
energy. The iteration is terminated using the discrete counterpart of
the stopping criterion \eqref{stopping-time}.


\section{Numerical simulation of coalescence}
\label{sec:simulations}

To test the sharp--diffuse algorithm, we revisit the symmetric
four-circle problem of Zhu--Chen--Hou~\cite[Fig.~14]{ZhChHo1996}.
The initial interface consists of two circles of radius $1$ centered at
$(\pm1.25,0)$ and two circles of radius $0.9$ centered at
$(0,\pm2)$. Under the nonlocal Mullins--Sekerka evolution, phase mass
is transferred between components, causing the
smaller components to contract while the larger components expand and approach one another. 
The calculation reported in \cite{ZhChHo1996} was halted at $t=0.75$
because the interface velocities became 
increasingly large near the coalescence.\footnote{%
It was observed in \cite{ZhChHo1996} that the impending coalescence evolves on a much
smaller time scale than the macroscale evolution, foreshadowing the
fast time scale \eqref{mixing-scale} of the localized mixing dynamics.
}

We continue the calculation through coalescence using the hybrid
sharp--diffuse algorithm. The formulation used in \cite{ZhChHo1996} is related to ours by
\begin{equation}\label{ZCH-scaling}
\tilde t
=
\frac{m_0\sigma}{4}t,
\qquad
\tilde\mu_0
=
\frac{2}{\sigma}\mu_0,
\qquad
\tilde\mu^{\mathrm f}
=
\frac{2}{\sigma}\mu^{\mathrm f},
\qquad
\widetilde V
=
\frac{4}{m_0\sigma}V, 
\end{equation}
where the tilde variables denote those used in \cite{ZhChHo1996}. 
Taking $m_0=\sigma=2$ gives
$\tilde t=t$, $\tilde\mu_0=\mu_0$, and $\tilde V=V$, reducing
\eqref{BI-eqs} exactly to the formulation of Zhu--Chen--Hou.

\begin{figure}[p]
\centering

\subfloat[Initial configuration, $t=0$]
{
\includegraphics[width=0.36\textwidth]
{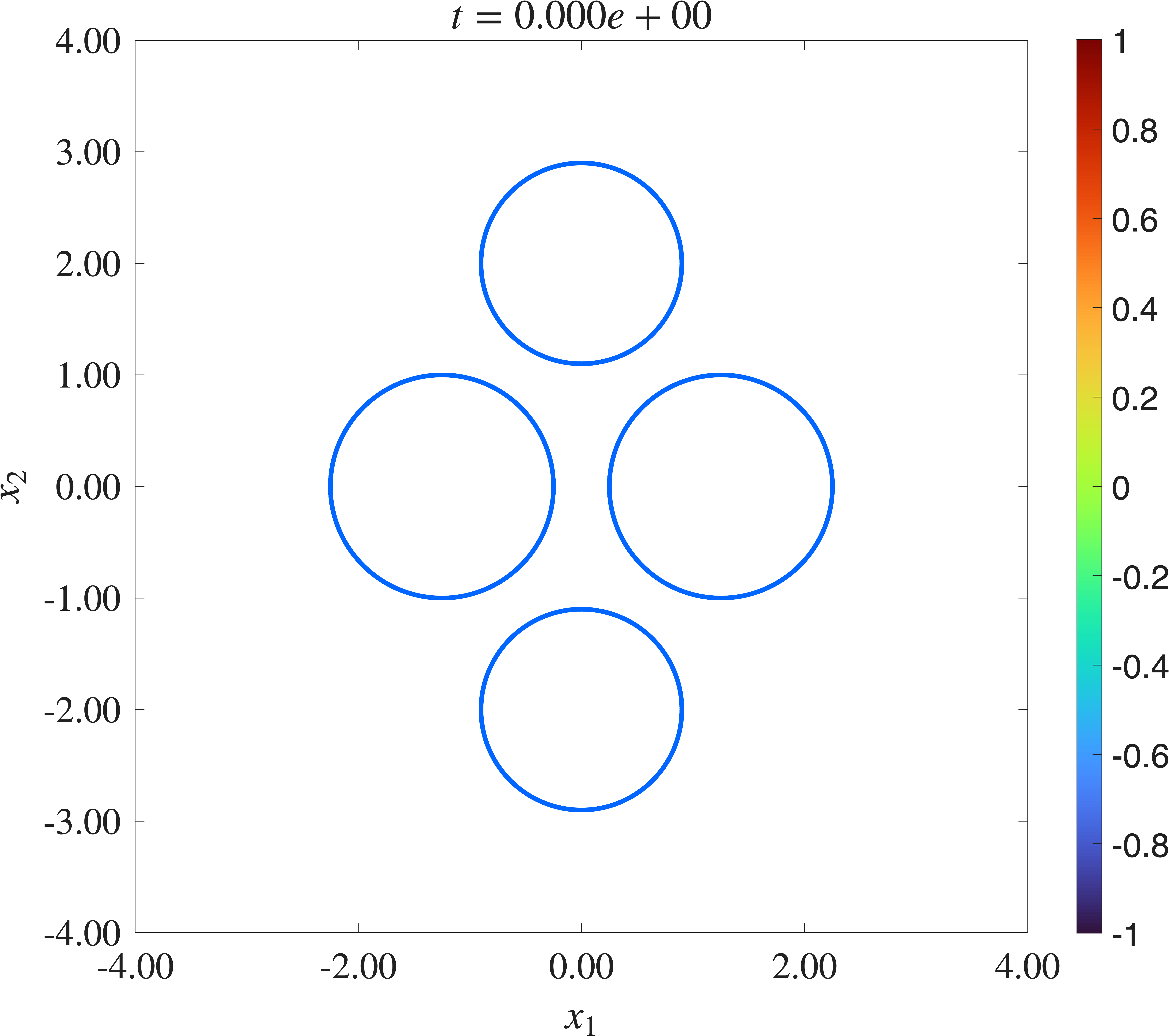}
}
\hspace{2em}
\subfloat[Incoming interface, $t=T^-$]
{
\includegraphics[width=0.36\textwidth]
{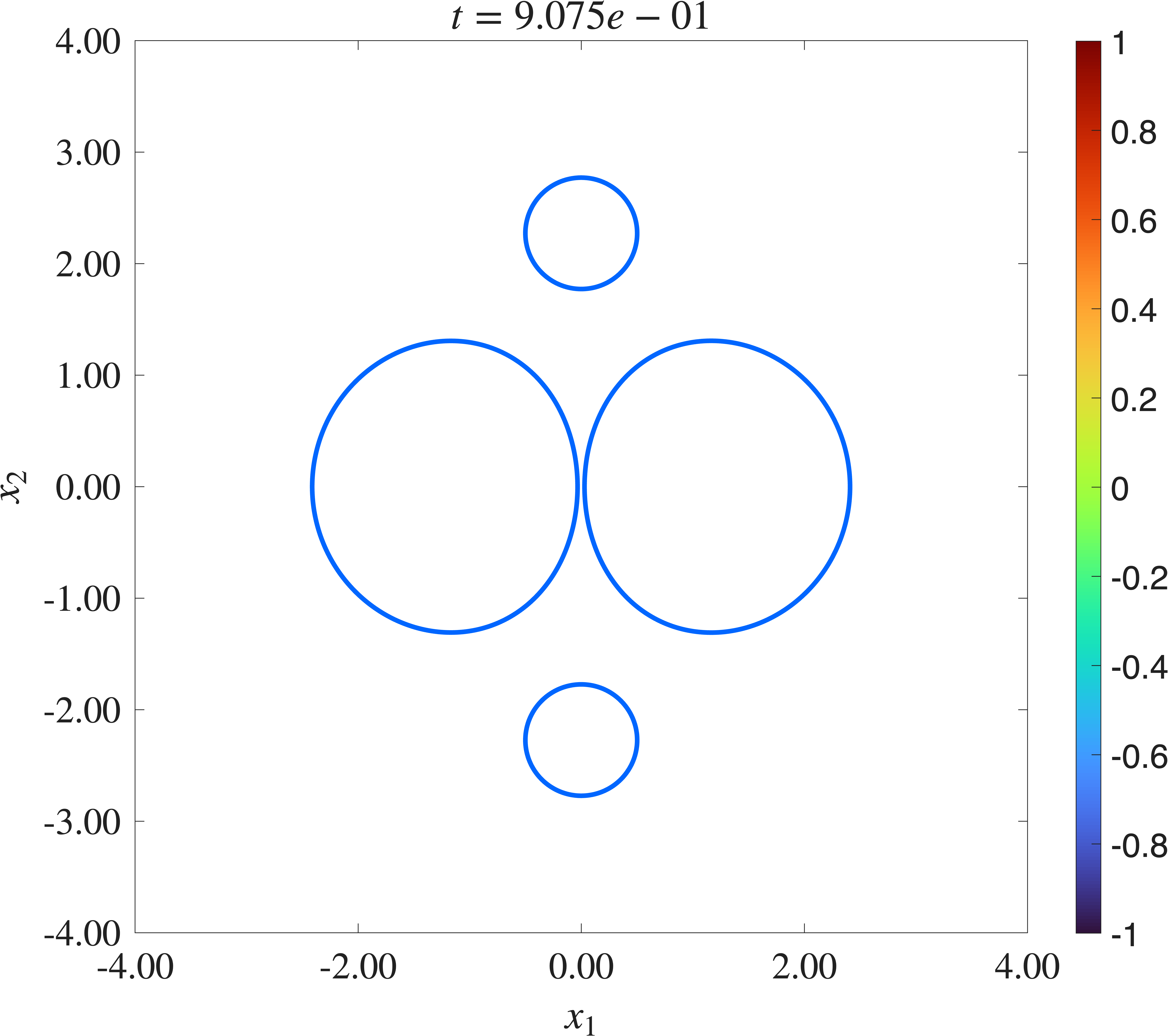}
}

\par\vspace{0.5em}

\subfloat[Initial phase field, $s=0$]
{
\includegraphics[width=0.37\textwidth]
{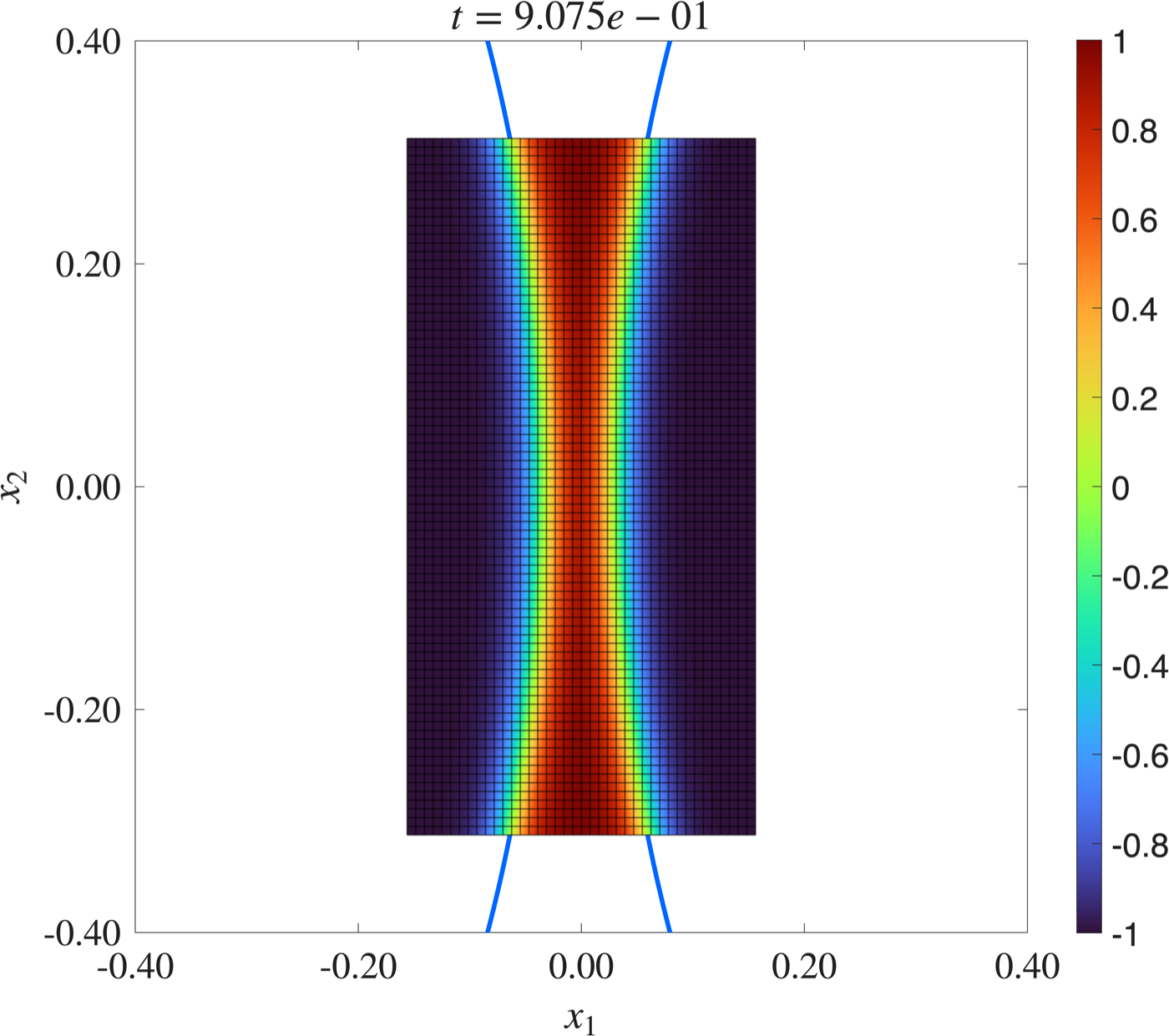}
}
\hspace{2em}
\subfloat[Relaxed phase field, $s=\smax$]
{
\includegraphics[width=0.37\textwidth]
{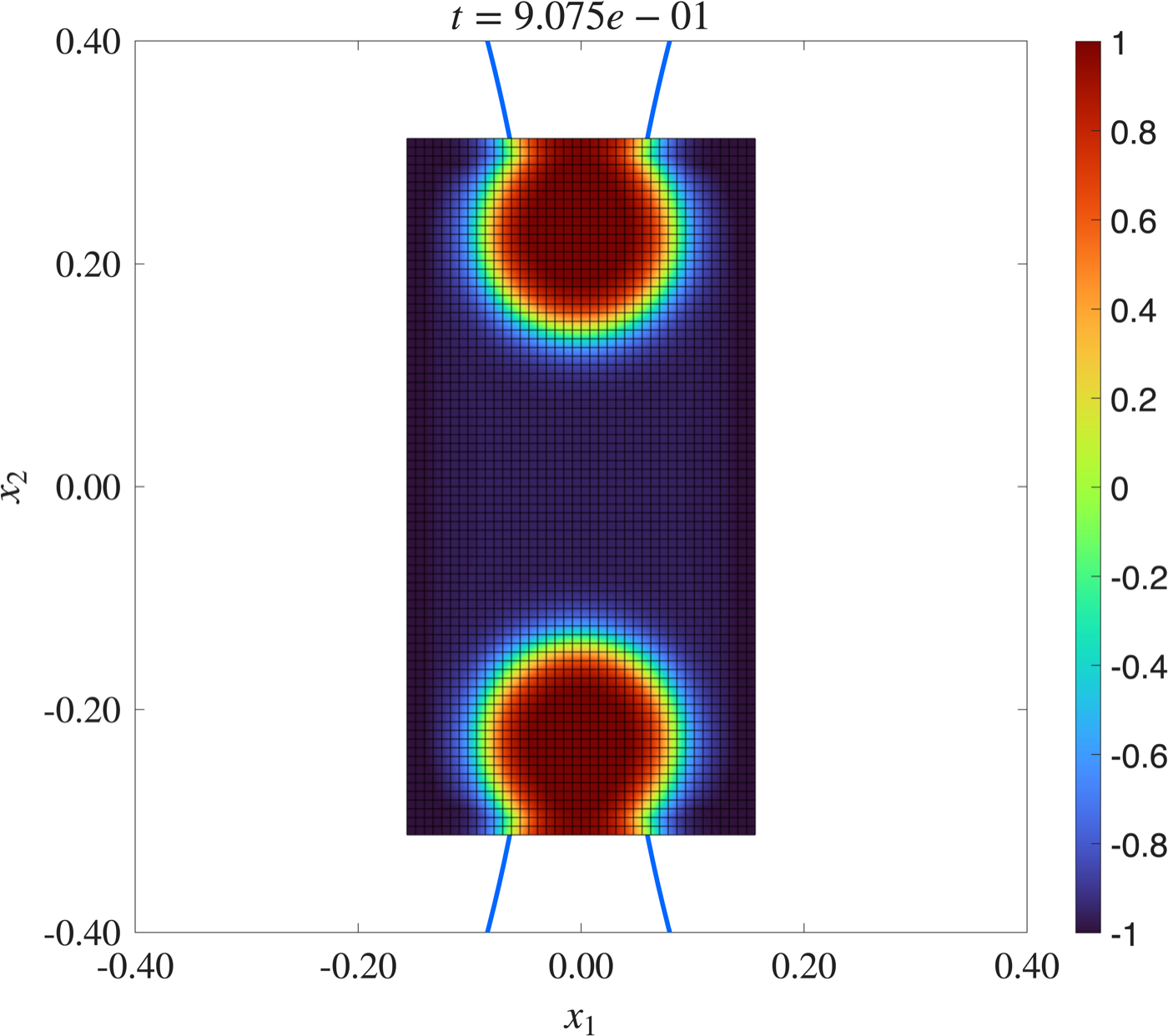}
}

\par\vspace{0.5em}

\subfloat[Outgoing interface, $t=T^+$]
{
\includegraphics[width=0.36\textwidth]
{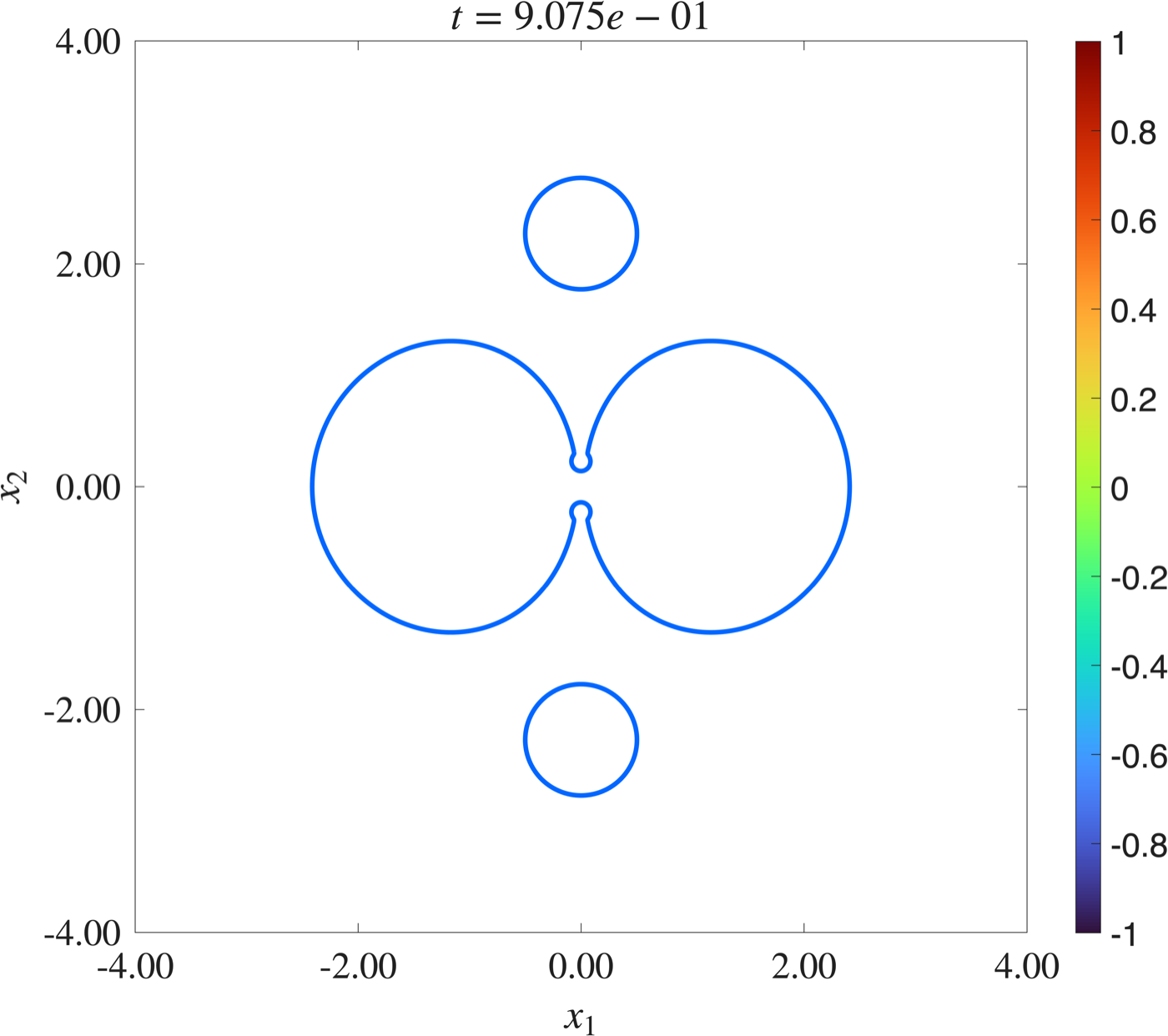}
}
\hspace{2em}
\subfloat[Restarted evolution, $t=1$]
{
\includegraphics[width=0.36\textwidth]
{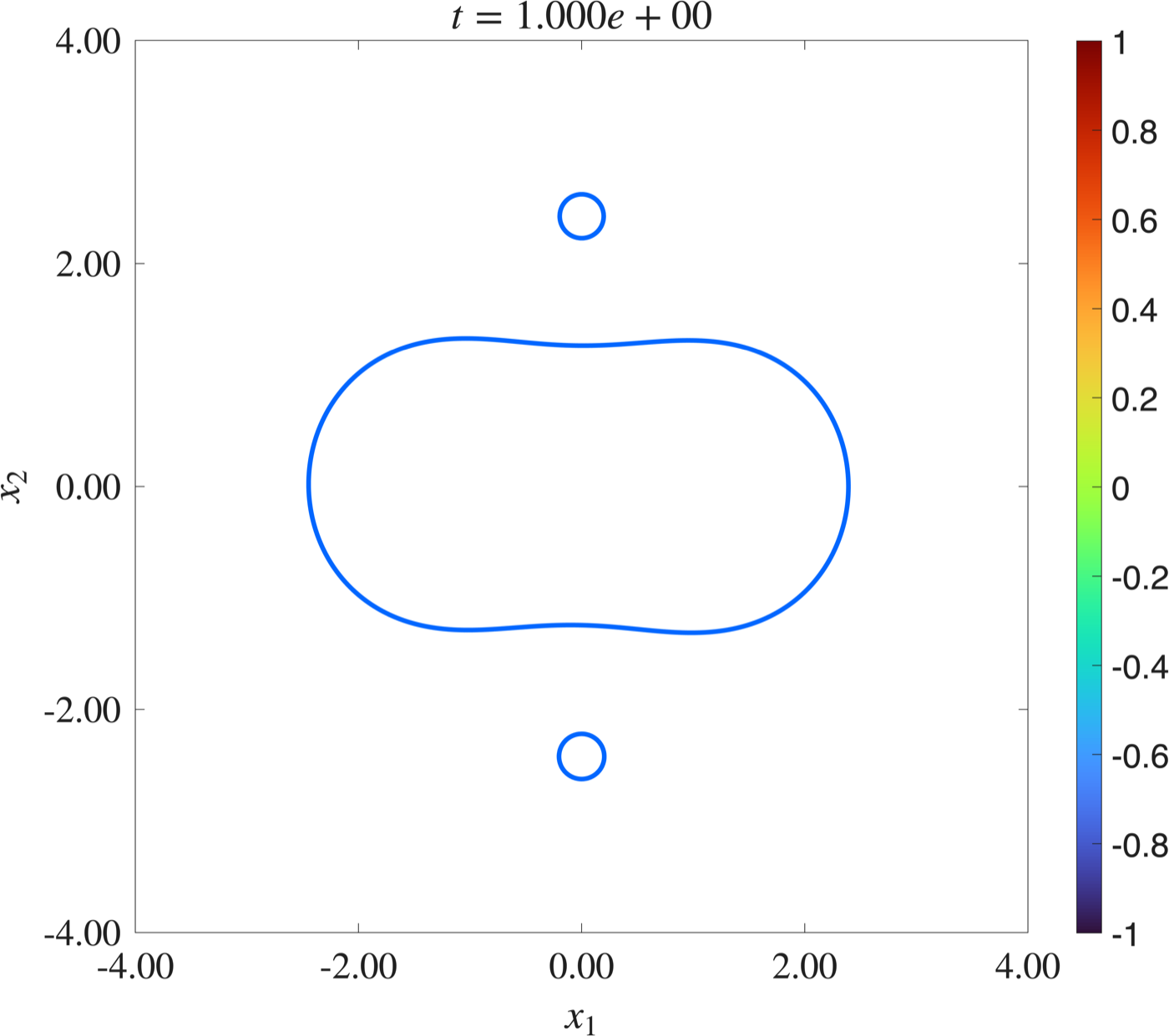}
}
\vspace{0.5em}
\caption{
\textbf{Simulation of coalescence using the sharp--diffuse interface
model.}
The four-circle problem of Zhu--Chen--Hou
\cite[Fig.~14]{ZhChHo1996} is simulated using
\Cref{alg:hybrid} with the parameters in
\eqref{four-circle-parameters} and
\eqref{four-circle-relaxation-parameters}.
The initial configuration in panel (a) evolves according to the
boundary-integral formulation (\Cref{prop:BI}) of the
Mullins--Sekerka system \eqref{MulSek}. While the transition layers
remain separated, \Cref{thm:sharp-limit} identifies this system as the
formal sharp-interface limit of the Cahn--Hilliard system \eqref{CH}.
As the two larger components approach one another, the geometric
separation required by \Cref{ass:separated-layers} degenerates.
The event detector in \Cref{alg:detector} halts the sharp-interface
evolution at the event time \eqref{four-circle-event-time}, shown in
panel (b). The corresponding sharp-interface calculation of Zhu--Chen--Hou was
terminated at $t=0.75$ prior to coalescence, and continued no further. 
Here, the localized phase-field system \eqref{localized-eqs}
is instead used to resolve the topological transition, as shown in panels (c)--(d);
during this relaxation, the macroscopic time $t$ is frozen, with $s$ denoting
pseudo-time. The terminal zero level set defines the outgoing
interface in panel (e), from which the sharp-interface evolution is restarted
and evolved until the final time $t=1$, as shown in panel (f). A video of the 
complete evolution (a)--(f) is available at \cite{RamaniWebsite}.
}
\label{fig:four-circle-evolution}
\end{figure}

Each interface is initialized with $32$
Lagrangian markers and adaptively refined so that the marker spacing
does not exceed $2\times10^{-2}$. This gives $316$ markers on each
larger circle and $284$ on each smaller circle. The principal
discretization parameters are
\begin{equation}\label{four-circle-parameters}
h
=
\frac{1}{128},
\qquad
\eps
=
2h
=
\frac{1}{64},
\qquad
\Delta t_{\mathsf{BI}}
=
2.5\times10^{-3}.
\end{equation}
Here $\Delta t_{\mathsf{BI}}$ is the maximum boundary-integral time
step, which is reduced as necessary to satisfy the interface-motion
CFL condition. The localized Cartesian patch contains
$40\times80$ grid cells. The event criterion is triggered at
\begin{equation}\label{four-circle-event-time}
T
=
0.9075.
\end{equation}
The sharp-interface evolution is then frozen while the localized
phase-field system is advanced in pseudo-time using
\begin{equation}\label{four-circle-relaxation-parameters}
\Delta s
=
10^{-4},
\qquad
\smax
=
1.41\times10^{-2}.
\end{equation}
After $141$ pseudo-time steps, the discrete relative
energy-dissipation rate falls below the tolerance in
\eqref{stopping-time}. The terminal zero level set defines the outgoing
interface, from which the boundary-integral calculation is restarted. 
Over $0\leq t\leq1$, the relative error in the conserved total
enclosed area is $0.27\%$.
The complete evolution is shown in
\Cref{fig:four-circle-evolution}.

For comparison, we compute the fully diffuse Cahn--Hilliard evolution
on the periodic square $[-4,4]^2$ using the SAV--BDF2 Fourier
pseudospectral method of \cite{ShXuYa2018}. The domain is discretized
on an $\Ngrid\times \Ngrid$ uniform grid, with
\begin{equation}\label{four-circle-diffuse-grid-scaling}
h
=
\frac{8}{\Ngrid},
\qquad
\eps
=
2h
=
\frac{16}{\Ngrid}, 
\qquad 
\Ngrid=512, \  1024, \ 2048. 
\end{equation}
To compare the fully diffuse and hybrid solutions at $t=1$, we reconstruct a phase
field from the final sharp interface using the equilibrium profile \eqref{sdf-phi}.

\begin{figure}[p]
\centering

\makebox[0.33\textwidth][c]{\textbf{Diffuse}}
\hspace{2em}
\makebox[0.33\textwidth][c]{\textbf{Sharp--Diffuse}}

\vspace{0.5em}

\subfloat[$\Ngrid=512$, $\eps=1/32$]
{
\includegraphics[width=0.33\textwidth]
{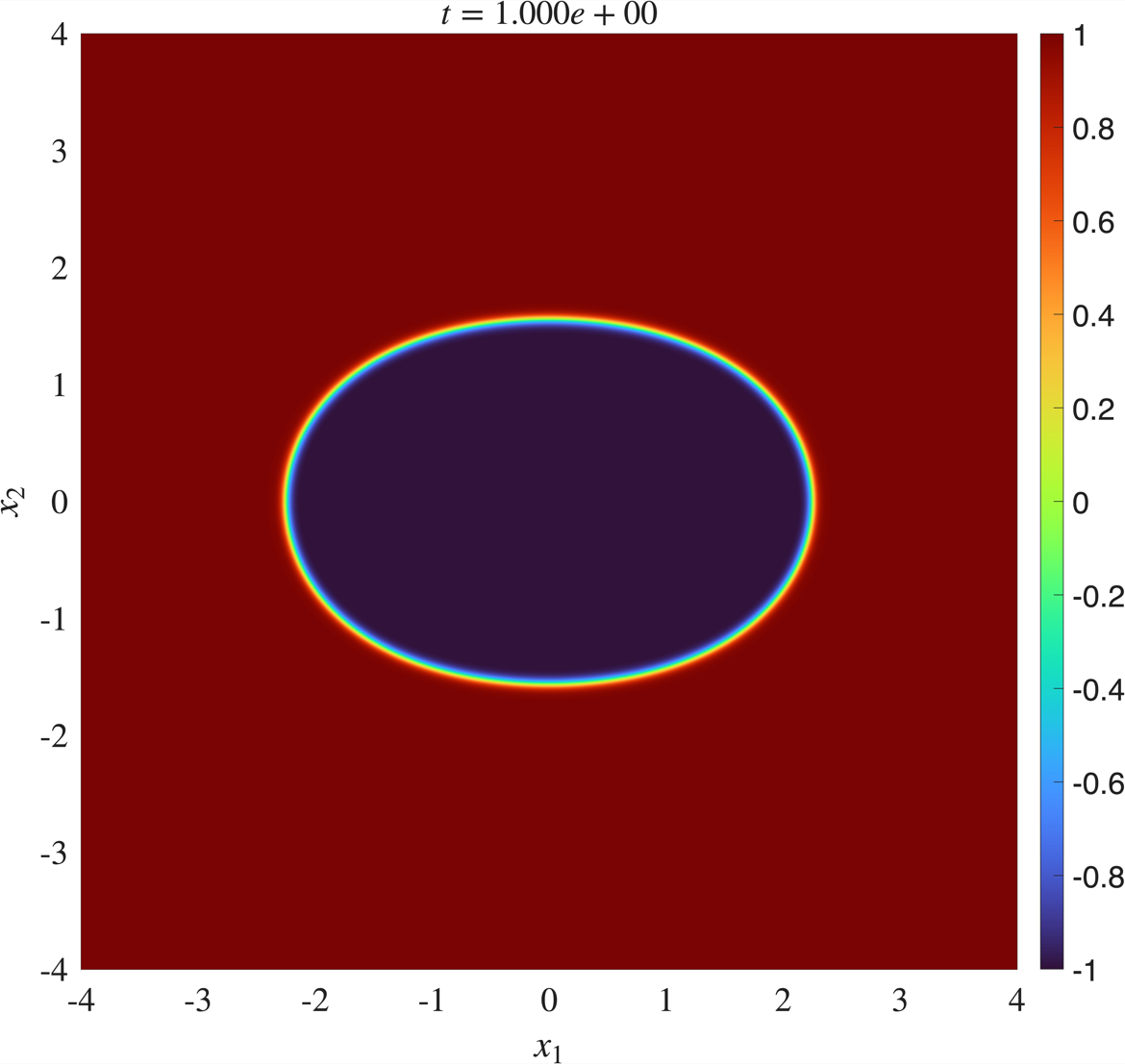}
}
\hspace{2em}
\subfloat[$\Ngrid=1024$, $\eps=1/64$]
{
\includegraphics[width=0.33\textwidth]
{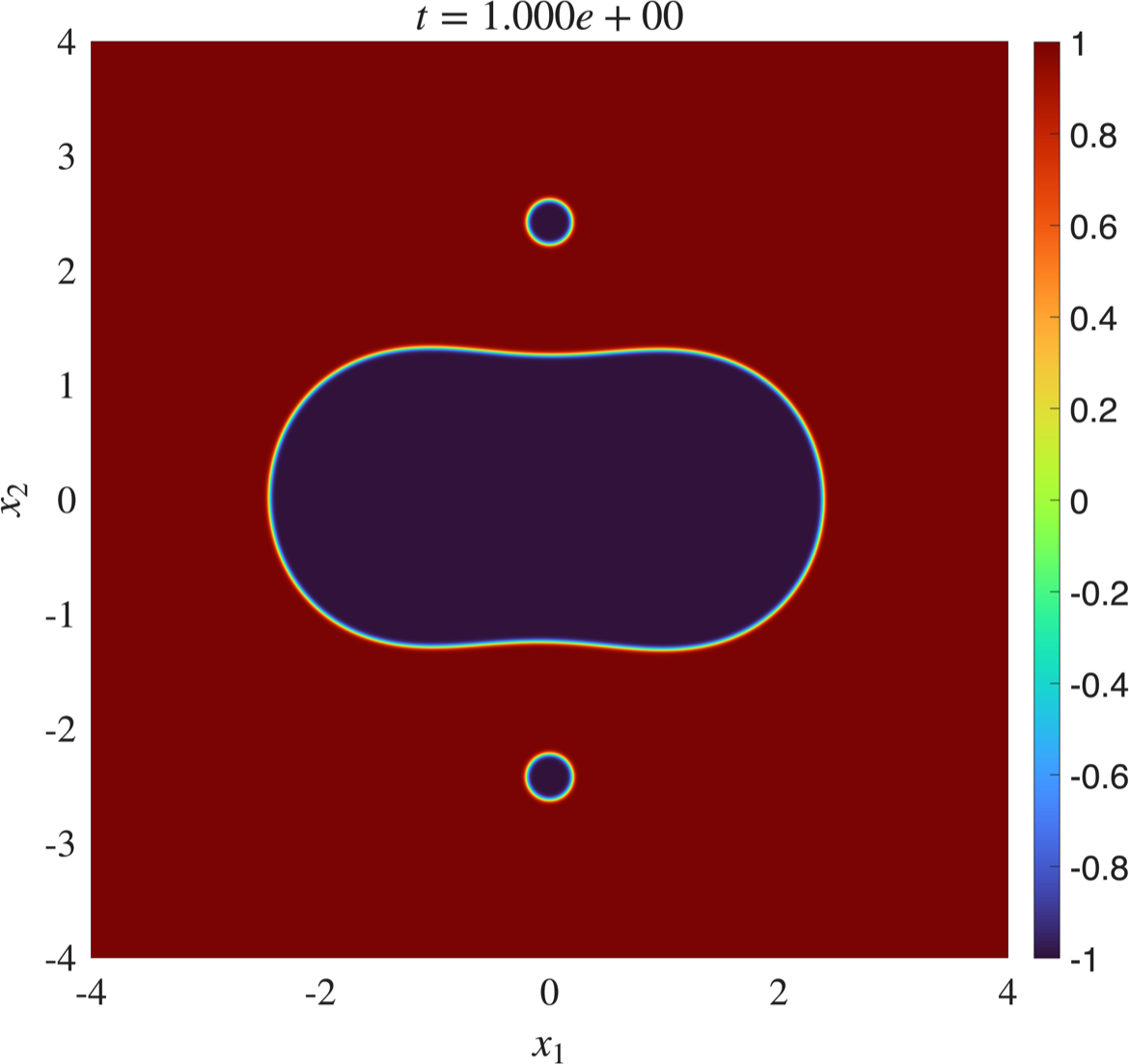}
}

\par\vspace{1em}

\subfloat[$\Ngrid=1024$, $\eps=1/64$]
{
\includegraphics[width=0.33\textwidth]
{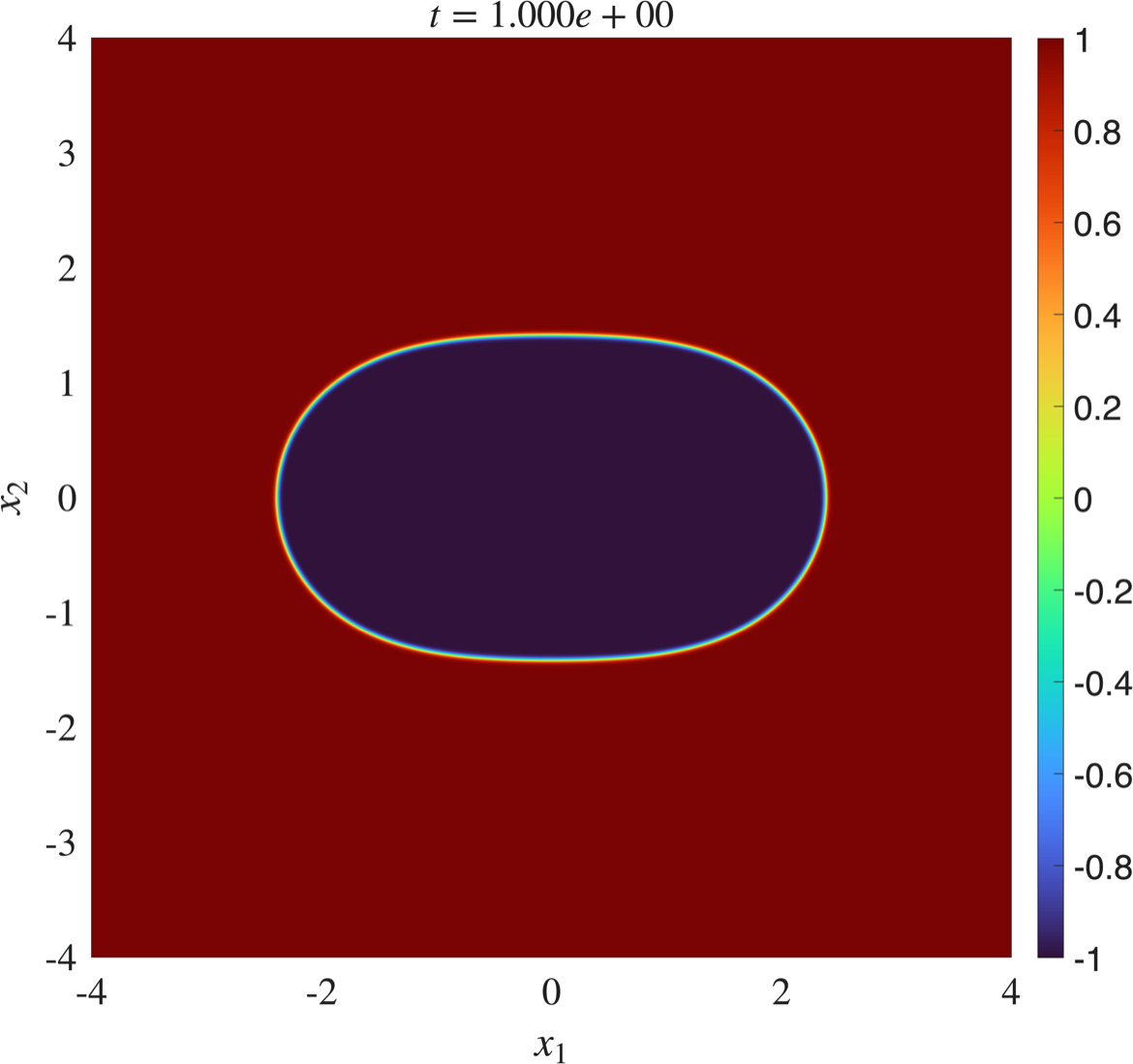}
}
\hspace{2em}
\subfloat[$\Ngrid=2048$, $\eps=1/128$]
{
\includegraphics[width=0.33\textwidth]
{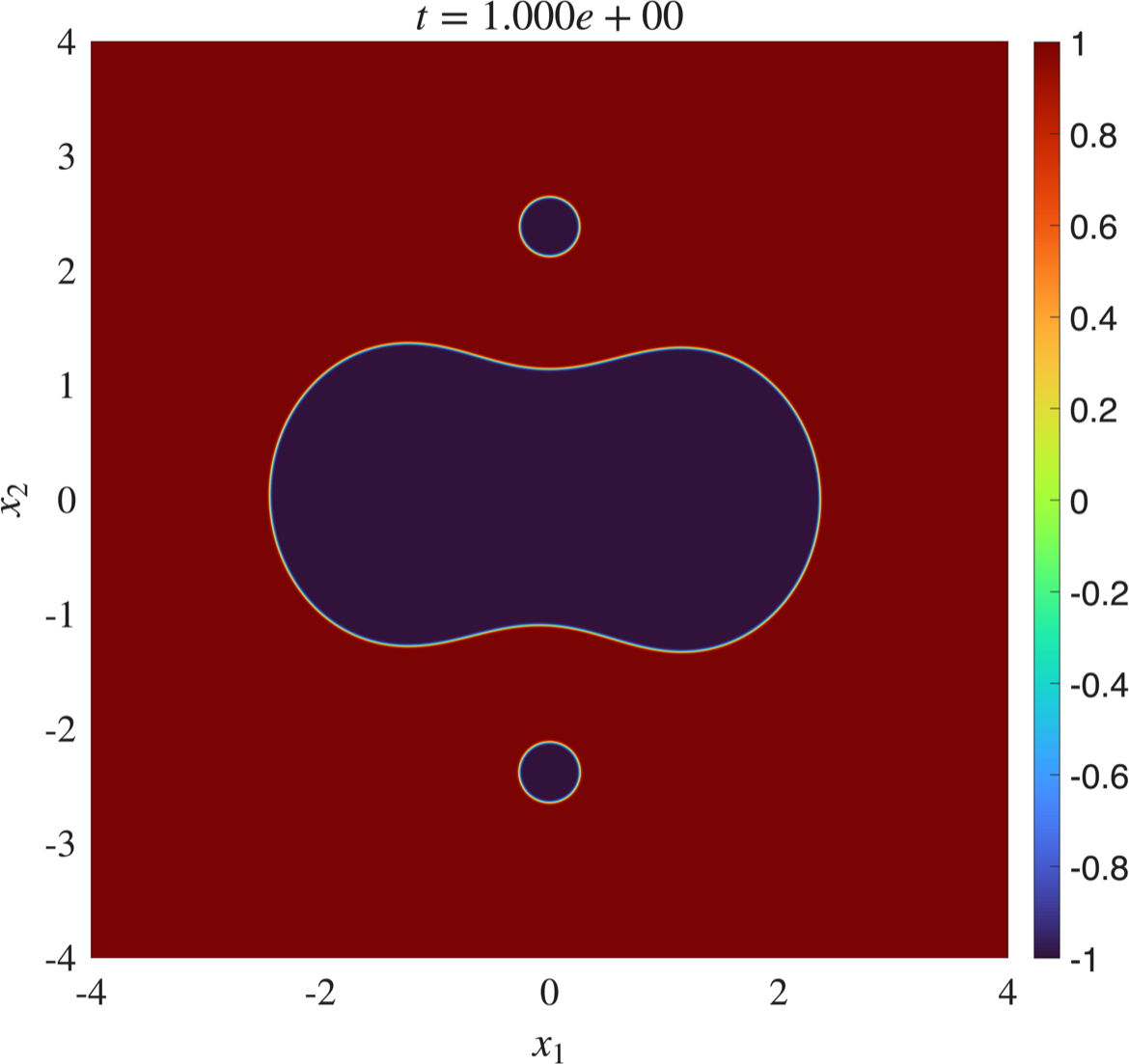}
}

\par\vspace{1em}

\subfloat[$\Ngrid=2048$, $\eps=1/128$]
{
\includegraphics[width=0.33\textwidth]
{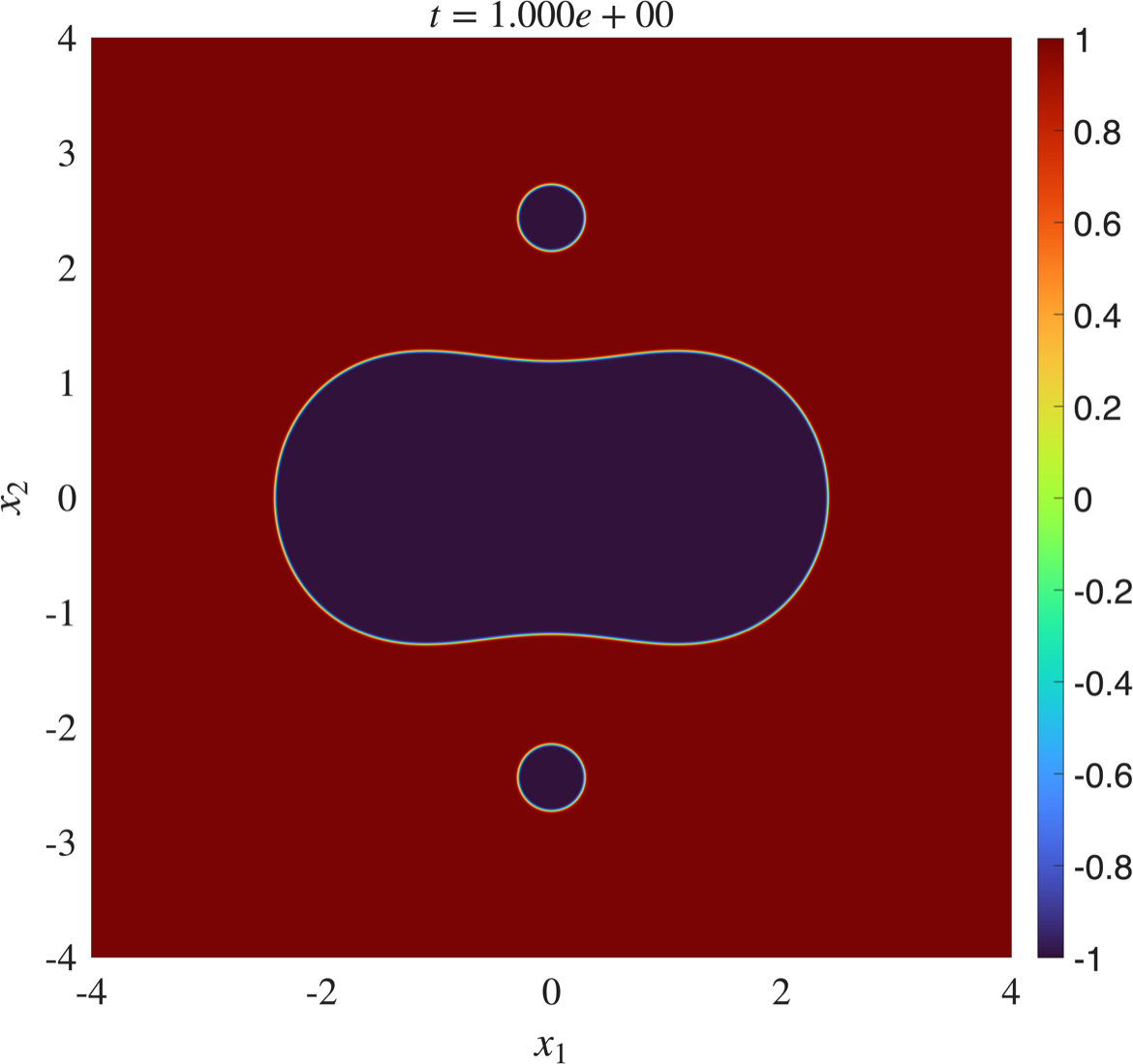}
}
\hspace{2em}
\subfloat[$\Ngrid=4096$, $\eps=1/256$]
{
\includegraphics[width=0.33\textwidth]
{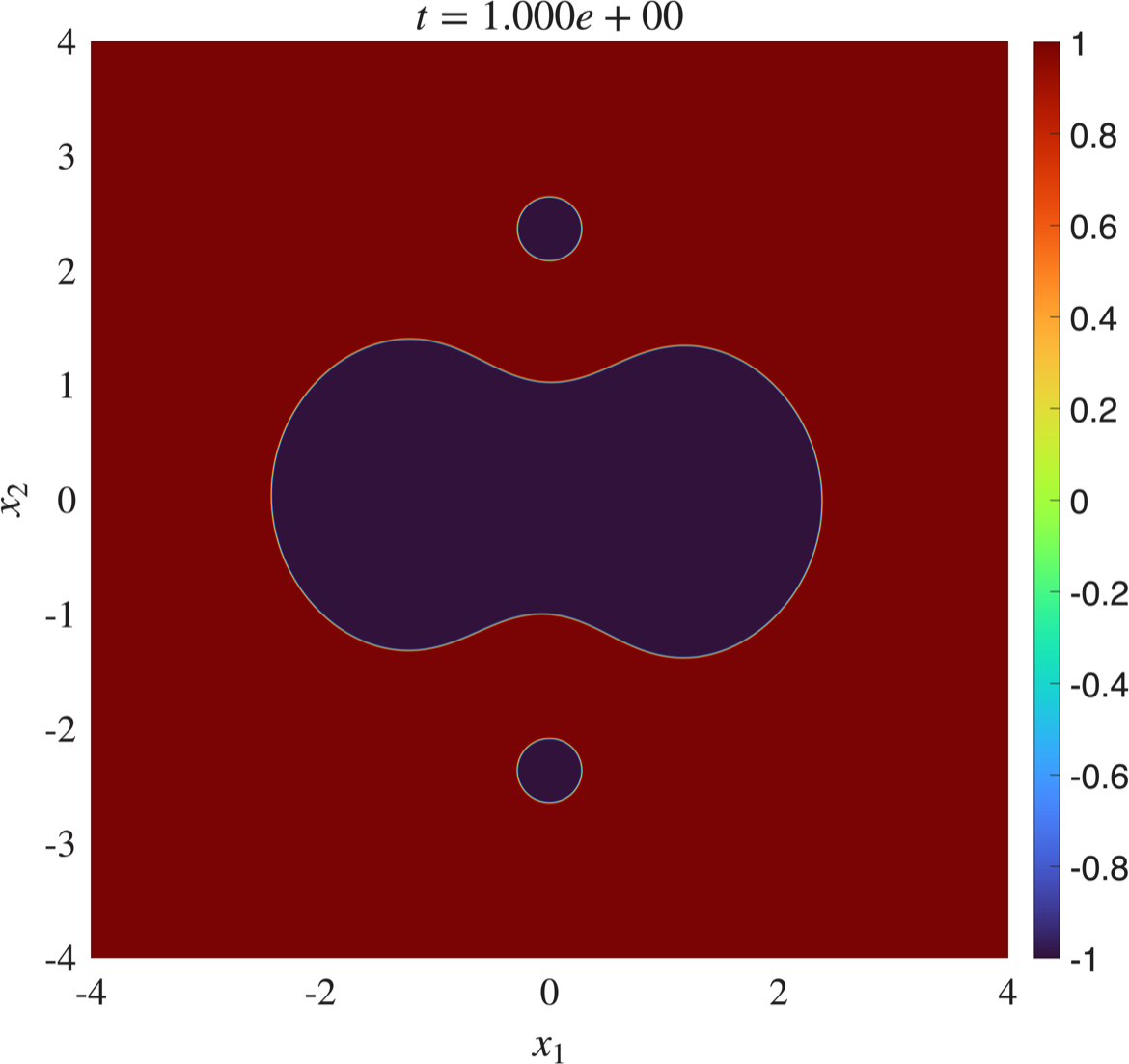}
}

\vspace{1em}
\caption{
\textbf{Comparison of diffuse and sharp--diffuse simulations
of coalescence as $\eps\to0$.}
The four-circle problem of Zhu--Chen--Hou
\cite[Fig.~14]{ZhChHo1996} is simulated using the fully diffuse
SAV--BDF2 Fourier pseudospectral method of \cite{ShXuYa2018} and the
sharp--diffuse interface model in \Cref{alg:hybrid}. 
The phase fields at $t=1$ are displayed, which show that the 
final morphology depends strongly on $\eps$. 
Larger $\eps$ produces earlier coalescence, leaving more time
for surface tension to round the connected interface and accelerate
the disappearance of the smaller components. Decreasing $\eps$ delays
both processes; in the fully diffuse calculation with
$\Ngrid=2048$, the smaller components persist at $t=1$ and
the coalesced component retains a dumbbell shape. The hybrid method
reproduces this qualitative small-$\eps$ behavior at substantially
lower cost. As reported in \Cref{tab:four-circle-runtimes}, at
$\Ngrid=2048$ it is approximately $386$ times faster than
the corresponding diffuse calculation.
}
\label{fig:four-circle-resolution-comparison}
\end{figure}

As shown in \Cref{fig:four-circle-resolution-comparison}, the final
morphology depends strongly on $\eps$. Larger values produce earlier
coalescence, leaving more time for interfacial relaxation and nonlocal
mass transfer to eliminate the smaller components. At
$\eps=1/128$, by contrast, the fully diffuse solution retains both
smaller components at $t=1$, while the central component remains
distinctly dumbbell-shaped. 
The sharp--diffuse method captures this qualitative small-$\eps$ behavior at
substantially lower cost. The computational runtimes reported in
\Cref{tab:four-circle-runtimes} show speedup factors of approximately $68$ and
$386$ at $\Ngrid=1024$ and $2048$, respectively. 
Meanwhile, at $\eps = 1/256$ and $\Ngrid = 4096$, the sharp--diffuse method simulates 
the evolution in just $86$ seconds, whereas the corresponding fully diffuse calculation 
is prohibitively expensive on the current computational platform. 

\begin{table}[ht]
\centering
\caption{
  \textbf{Wall-clock times for the four-circle coalescence problem.}
}
\label{tab:four-circle-runtimes}
\begin{tabular}{
  l@{\hskip 16pt}
  c@{\hskip 12pt}
  c@{\hskip 12pt}
  c
}
\toprule
Method
& $\Ngrid$
& $\eps$
& $T_{\mathsf{CPU}}$
\\
\midrule
\multirow{3}{*}{Diffuse}
& $512$  & $1/32$  & $4\,\mathrm{min}$  \\
& $1024$ & $1/64$  & $35\,\mathrm{min}$ \\
& $2048$ & $1/128$ & $4.5\,\mathrm{h}$   \\
\midrule
\multirow{3}{*}{Sharp--Diffuse}
& $1024$ & $1/64$  & $31\,\mathrm{s}$ \\
& $2048$ & $1/128$ & $42\,\mathrm{s}$ \\
& $4096$ & $1/256$ & $86\,\mathrm{s}$ \\
\bottomrule
\end{tabular}
\end{table}

Finally, we characterize the immediate post-coalescence dynamics by
computing the bridge radius, 
\begin{equation}\label{bridge-radius-definition}
\begin{aligned}
r_b(t)
&\coloneqq
\frac{x_2^+(t)-x_2^-(t)}{2},\\
x_2^+(t)
&\coloneqq
\min\{x_2\geq0:(0,x_2)\in\Gamma(t)\}, \\
x_2^-(t)
&\coloneqq
\max\{x_2\leq0:(0,x_2)\in\Gamma(t)\}.
\end{aligned}
\end{equation}
In studies of liquid-drop coalescence, the bridge radius is a standard
measure of the post-contact dynamics. At early times, it typically
obeys a self-similar law $r_b\sim t^\alpha$, with the exponent $\alpha$
determined by the geometry and dominant transport mechanism
\cite{EgSpSn2025,LiEgFoSp2026}.
In particular, the bulk diffusion considered here is predicted to produce the
exponent $\alpha=1/5$ \cite[Sec.~2.5]{EgSpSn2025}, a growth law that
has been observed experimentally during the sintering
of metal spheres \cite{Kuczynski1949}.

\begin{figure}[tbhp]
\centering

\subfloat[$\Ngrid=1024$]{%
    \includegraphics[width=0.3\textwidth]
    {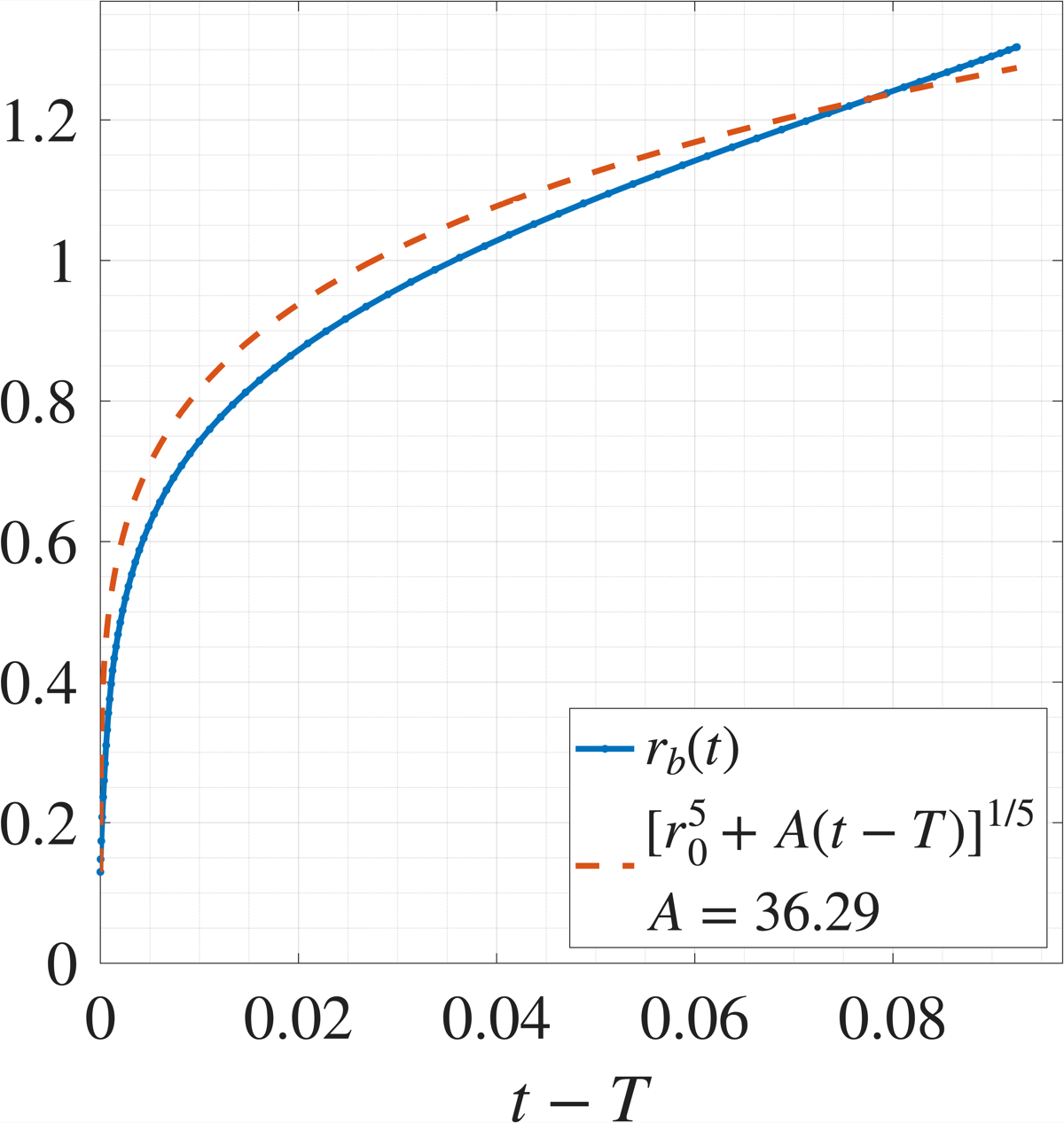}%
}
\hspace{1em}
\subfloat[$\Ngrid=2048$]{%
    \includegraphics[width=0.3\textwidth]
    {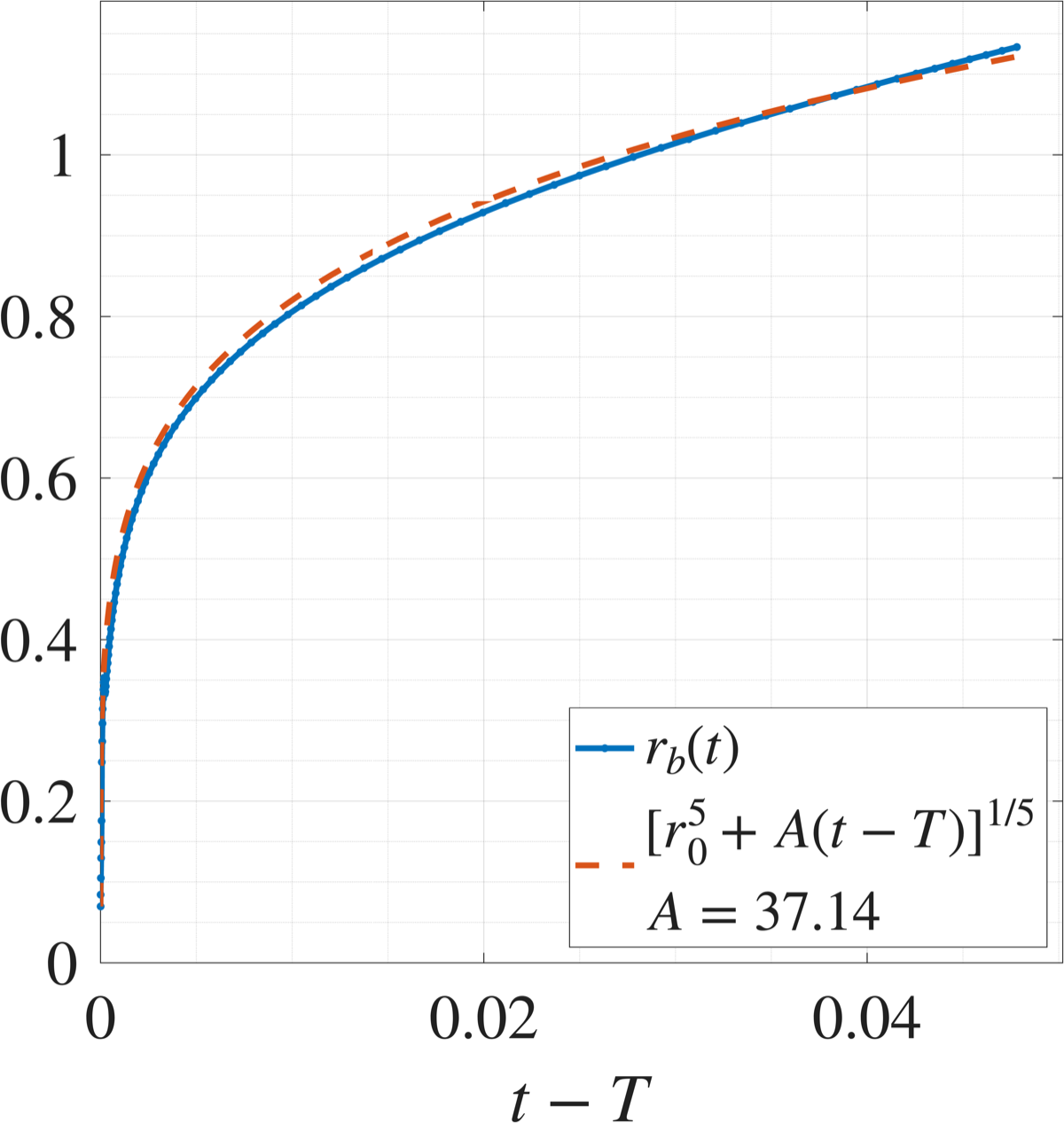}%
}
\hspace{1em}
\subfloat[$\Ngrid=4096$]{%
    \includegraphics[width=0.3\textwidth]
    {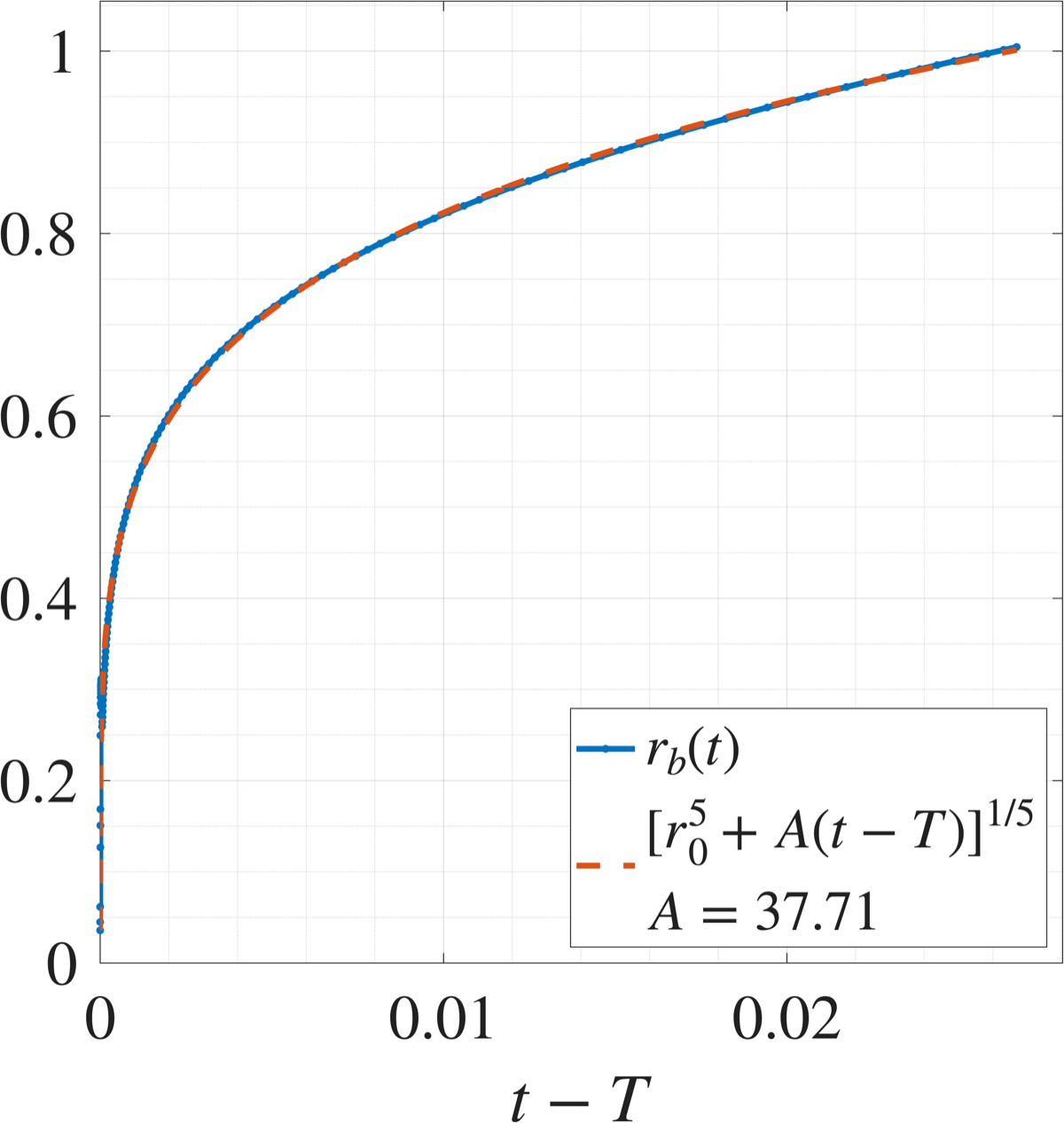}%
}

\vspace{0.5em}
\caption{
\textbf{Self-similar bridge growth following coalescence.}
The bridge radius \eqref{bridge-radius-definition} is shown for
three sharp--diffuse simulations. The dashed curves are least-squares
fits to the predicted law \eqref{bridge-growth-law} over
$[T,\tmax]$. Agreement with the predicted scaling improves as the
resolution is increased.
}
\label{fig:bridge-growth}
\end{figure}

Specifically, if $R$ is the macroscopic radius
of curvature of the incoming interfaces, then the characteristic axial half-width
satisfies $w\sim r_b^2/R$ for $r_b \ll R$. 
Over this axial scale, the interface undergoes a radial displacement
of order $r_b$, so the radius of curvature at the bridge tip satisfies
$r_c\sim w^2/r_b\sim r_b^3/R^2$, and hence
$\kappa_b\sim R^2/r_b^3$.
The Gibbs--Thomson condition \eqref{GibbsThomson} then gives the
chemical-potential scale
$|\mu_0|\sim\sigma R^2/r_b^3$.
Assuming that the corresponding harmonic field varies over the radial
bridge scale $r_b$, its normal derivative jump satisfies
$|[\p_n\mu_0]|\sim|\mu_0|/r_b\sim\sigma R^2/r_b^4$.
Substituting into the velocity law
\eqref{MS-velocity} and integrating yields the self-similar law
\begin{equation}\label{bridge-growth-law}
r_b(t)
= 
\left[r_0^5+A(t-T)\right]^{1/5},
\qquad t\geq T, 
\end{equation}
where $T$ is the topological event time, $r_0=r_b(T^+)$ is the initial
post-transition bridge radius, and $A>0$ is a constant which can be 
approximated by least-squares fitting over the time interval $[T,\tmax]$.
We compute the bridge growth for the sharp--diffuse simulations with
$\Ngrid=1024$, $2048$, and $4096$, for which the event times are
$T=0.9075$, $0.9522$, and $0.9733$, respectively.
The results in \Cref{fig:bridge-growth} show increasingly close
agreement with the predicted law \eqref{bridge-growth-law} 
as the resolution is increased.

\section{Conclusions}
\label{sec:conclusions}

\begin{figure}[htbp]
\centering

\includegraphics[width=0.60\textwidth]
{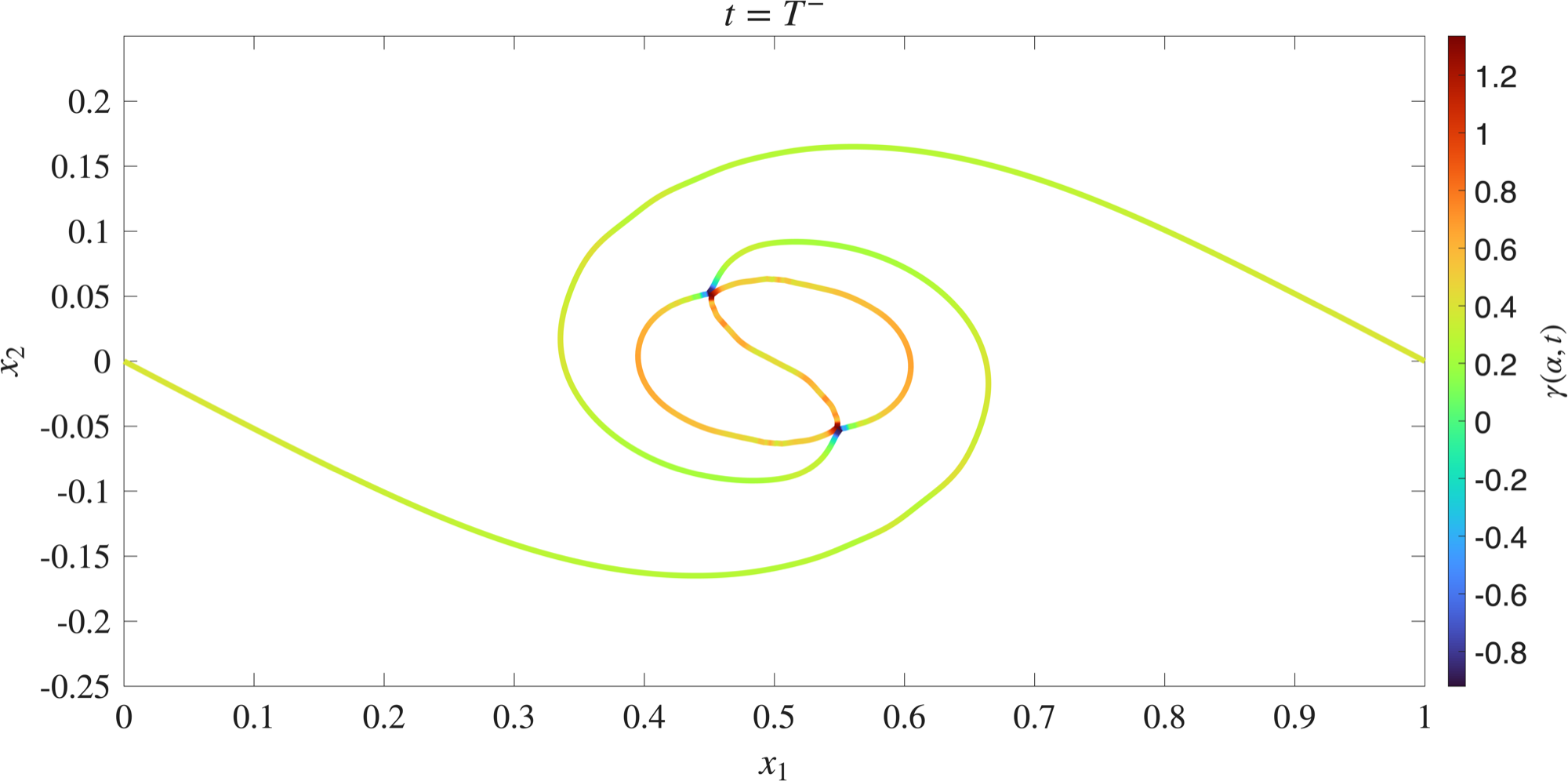}

\par\vspace{0.5em}

\includegraphics[width=0.60\textwidth]
{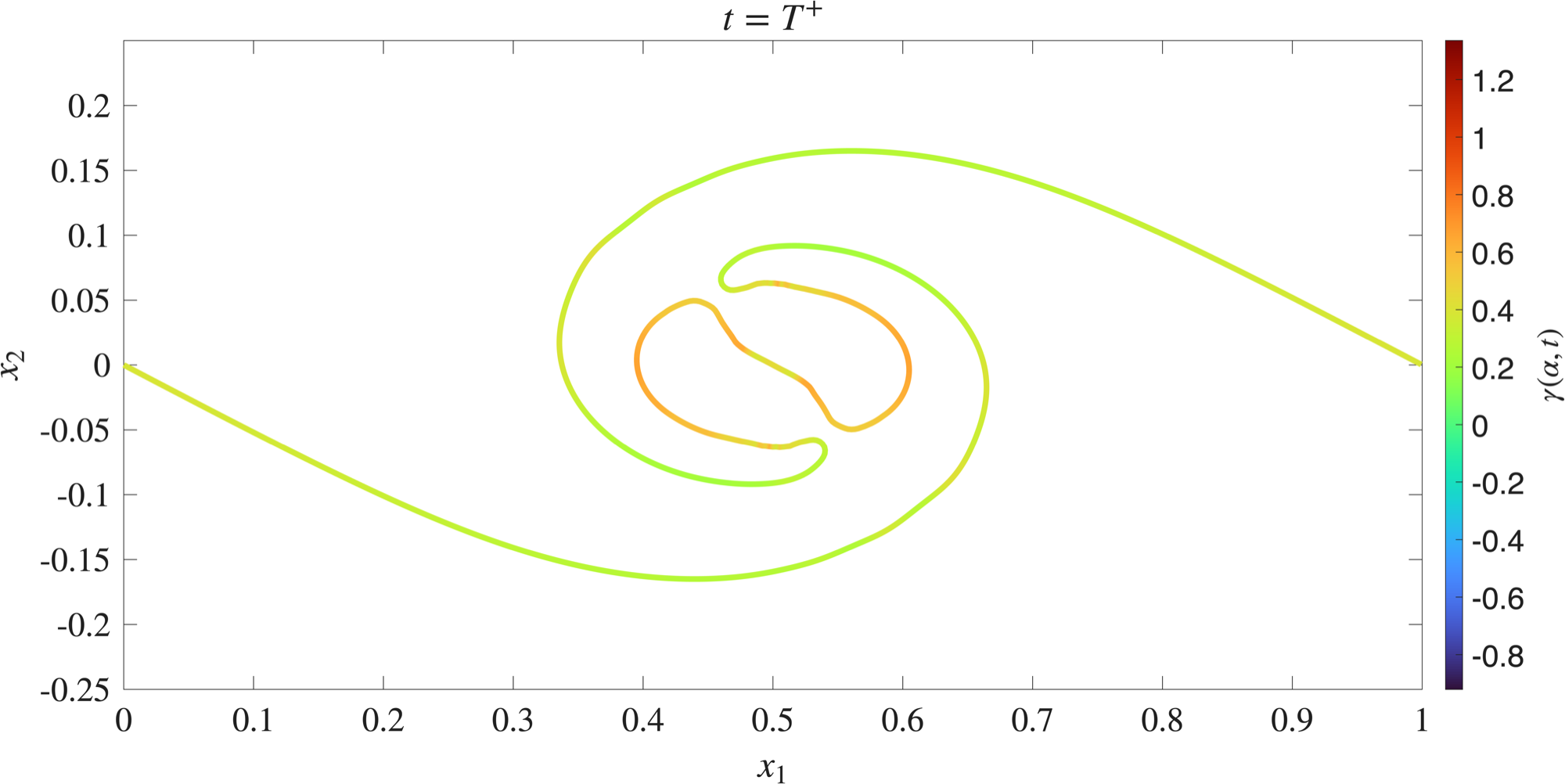}

\par\vspace{0.5em}

\includegraphics[width=0.60\textwidth]
{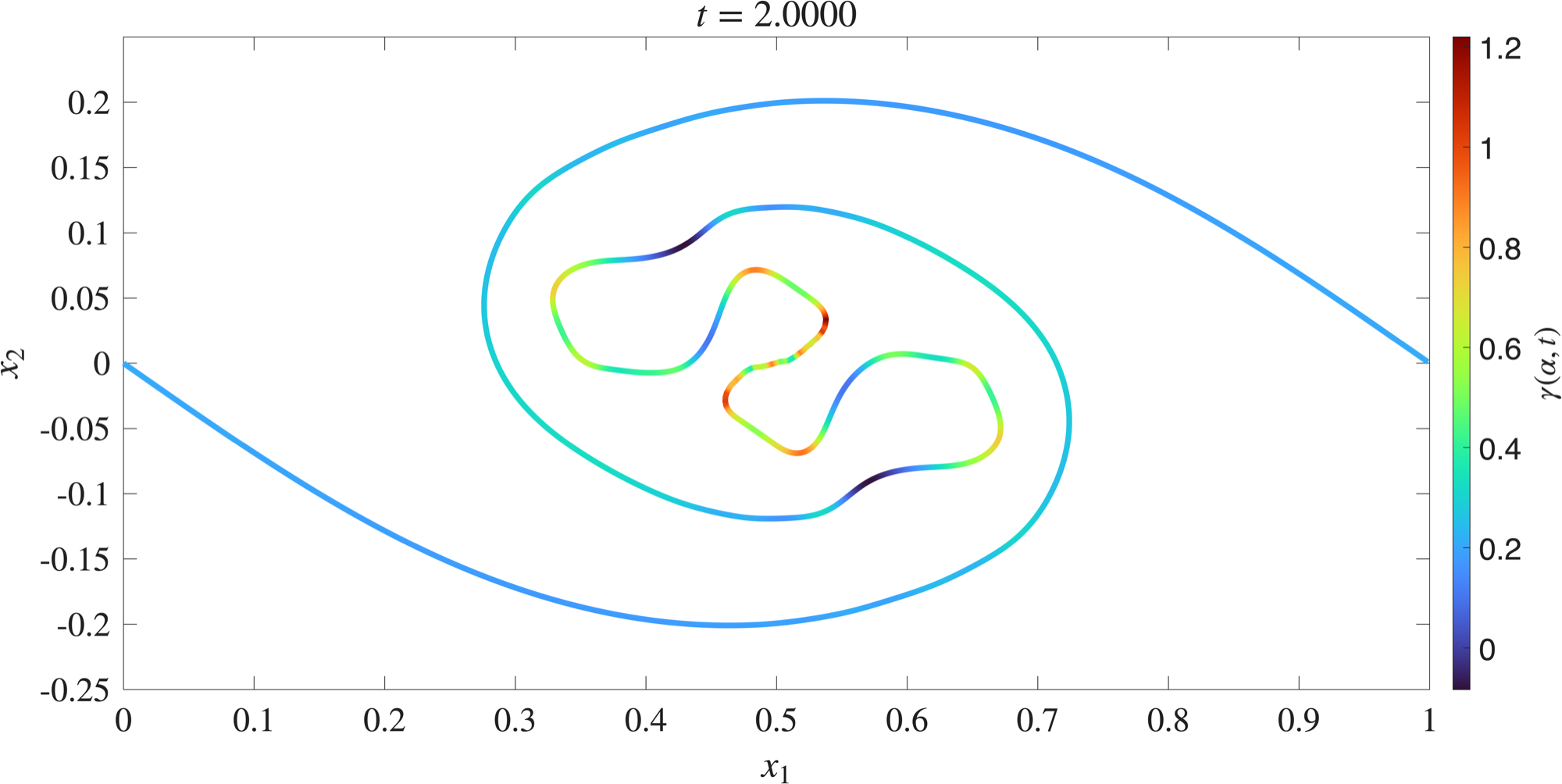}

\caption{
\textbf{Sharp--diffuse continuation through vortex-sheet pinch-off.}
We consider the periodic vortex-sheet problem
of Hou--Lowengrub--Shelley~\cite[Figure~11]{HoLoSh1994}, in which a
sinusoidal perturbation of a flat sheet rolls up under the
Kelvin--Helmholtz instability in a homogeneous fluid at Weber number
$\mathrm{We}=200$.
This preliminary computation uses $N=1024$ interface points and
transition-layer width $\eps=2h=1/512$, where $h=1/1024$ is the
local grid spacing. The interface is shown at the topological event
time $t=T^-$ (top), following localized diffuse relaxation at
$t=T^+$ (middle), and after restarting the sharp-interface evolution
at the final time $t=2$ (bottom). The line color denotes the
vortex-sheet strength $\gamma$. The topological event is detected at
$T=1.5$, whereas the computations in \cite{HoLoSh1994} terminated at $t=1.4$.
The symmetry of the configuration produces two simultaneous
topological events; the sharp--diffuse solution preserves this
symmetry through the topology change.
}
\label{fig:vortex-sheet-evolution}
\end{figure}

We have developed a sharp--diffuse interface model for intermittent
and isolated topological transitions in Cahn--Hilliard phase coarsening.  The
mathematical justification for the model is based on a geometric characterization of
the sharp-interface regime and established analytical results for the
limiting Mullins--Sekerka system.  The model couples efficient
boundary-integral evolution with spacetime-localized diffuse
relaxation.  Numerical tests demonstrate continuation through
coalescence and speedups of two to three orders of magnitude over a fully
diffuse solver. In the future, we plan to extend the model to more complex interfacial
flows. A preliminary sharp--diffuse computation for a vortex sheet with surface
tension is shown in \Cref{fig:vortex-sheet-evolution}, demonstrating
continuation through the \emph{pinching} singularities observed by 
Hou--Lowengrub--Shelley~\cite[Figure~11]{HoLoSh1994}.

Several analytical questions remain open.  Although the classical
Cahn--Hilliard and Mullins--Sekerka stages rest on established theory,
the interface-surgery construction remains formal and is justified only
for an idealized coalescence geometry.  For perturbations of this geometry, an 
implicit-function argument could establish both $C^{3+\alpha}$ regularity of the 
outgoing interface and the required lower bound on its reach. 
From the numerical point of view, simulating 
more complex phase-coarsening problems, particularly in
three dimensions, will require a robust, parallel implementation
with an event-local hierarchical data structure \cite{Ramani2026}.

Finally, we remark that the proposed methodology may be placed in the broader 
context of methods for continuation through singularities. 
The current program seeks continuation through topological singularities; an 
analogous program is development of shock singularities for the
multidimensional compressible Euler equations.
In both settings, weak solutions do not select
a sufficiently resolved continuation, motivating two stages:
classical evolution followed by strong continuation.
Here, Mullins--Sekerka evolution is followed by
localized Cahn--Hilliard dynamics.  In the shock development program, 
classical compressible Euler evolution is followed
by free-boundary shock development; see \cite{Christodoulou2007,ShVi2024} for the
analysis of the shock formation stage, and \cite{Christodoulou2019,BuDrShVi2022} for the 
shock development stage.

The analogy between the two programs, summarized in \Cref{tab:analogy}, 
is architectural: a geometric criterion degenerates,
and a second problem constructs the strong continuation.
Two differences are essential.  First, the present
program is algorithmic and partly formal, whereas
the cited shock constructions are rigorous.
Second, our topological events are intermittent and isolated,
after which a smooth interface re-emerges.  By contrast,
the initial shock formation is followed instantaneously
by a continuum of singularities (i.e., shock propagation), while
weaker cusp and characteristic singularities also emanate from the first singularity,
further complicating the problem.

\begin{table}[t]
\centering
\caption{
\textbf{Structural analogy between sharp--diffuse continuation and shock
development.}
}
\label{tab:analogy}
\scriptsize
\renewcommand{\arraystretch}{1.35}
\begin{tabular}{
@{}
>{\raggedright\arraybackslash}p{0.22\textwidth}
>{\raggedright\arraybackslash}p{0.28\textwidth}
>{\raggedright\arraybackslash}p{0.28\textwidth}
@{}
}
\toprule
&
\textbf{Sharp--Diffuse}
&
\textbf{Shock development}
\\
\midrule
Regular evolution
&
Mullins--Sekerka
&
Smooth compressible Euler
\\
Singularity
&
Topological transition
&
Shock formation
\\
Tracked interfaces
&
Material
&
Fast acoustic characteristic
\\
Physical breakdown
&
Overlap of diffuse transition layers
&
Collapse of acoustic characteristics
\\
Weak continuation
&
Varifold solution
&
Convex integration
\\
Strong continuation
&
Localized Cahn--Hilliard evolution
&
Free-boundary shock development
\\
Outcome
&
Smooth interfaces satisfying the separated-layer condition
&
A persistent shock separating smooth flow regions
\\
\bottomrule
\end{tabular}
\end{table}


\appendix

\section{Polynomial profile for idealized coalescence} 
\label{app:polynomial-profile}

We derive the explicit polynomial height function
stated in Example~\ref{ex:ideal-polynomial-profile}. For prescribed values of
$L$, $H$, and $h(0)$, we show that it is the unique even polynomial of
degree at most twelve satisfying the attachment and area constraints in
Assumption~\ref{ass:ideal-coalescence}. We then verify that
\eqref{ideal-polynomial-admissibility} guarantees positivity and a
nondegenerate minimum. The polynomial was derived with the
assistance of ChatGPT and independently verified by direct substitution.

\subsection{Polynomial ansatz}

Recall from Example~\ref{ex:ideal-polynomial-profile} the dimensionless
parameters $\Lambda$ and $\alpha$, the symmetric variable $r$, and the
representation \eqref{ideal-polynomial-h}. The conditions
\eqref{ideal-reach-compatible-scales} and
\eqref{ideal-centre-conditions} imply that
$\Lambda>1$ and $0<\alpha<\Clyr/H<1$.

Since $r$ depends only on $\xi_1^2$, the representation
\eqref{ideal-polynomial-h} enforces evenness automatically. 
Conversely, every even polynomial $h$ of degree at most twelve admits a unique
representation of this form with $\mathcal{P}$ a polynomial of degree
at most six. The attachment and central-height conditions become
\begin{equation}\label{poly-end-values}
\mathcal{P}(0)=1,
\qquad
\mathcal{P}(1)=\alpha.
\end{equation}
Finally, introducing $s=\xi_1/L$, the area
constraint \eqref{ideal-mass-constraint} becomes
\begin{equation}\label{poly-area}
\int_0^1
\mathcal{P}(1-s^2)\dd{s}
=
\frac{\pi}{4\Lambda}.
\end{equation}

\subsection{Imposing the attachment conditions}

Since $\mathcal{P}$ has degree at most six and
$\mathcal{P}(0)=1$, we may write
\begin{equation}\label{poly-general-form}
\mathcal{P}(r)
=
1+\sum_{\ell=1}^6 c_\ell r^\ell.
\end{equation}
The derivative attachment conditions
\eqref{ideal-orthogonal-contact} and
\eqref{ideal-boundary-conditions}
determine the first four coefficients. By evenness, it suffices to
impose these conditions at $\xi_1=L$.
Rather than differentiate the composite expression
$h(\xi_1)=H\mathcal{P}(1-\xi_1^2/L^2)$ four times, we compare its Taylor
expansion at the attachment point. Set
\begin{equation}\label{poly-delta}
\delta
\coloneqq
\frac{\xi_1-L}{L},
\qquad
s=1+\delta,
\qquad
r=-2\delta-\delta^2.
\end{equation}
It follows that
\begin{equation}\label{poly-composition}
\mathcal{P}(-2\delta-\delta^2)
=
\frac{h(L+L\delta)}{H}.
\end{equation}
Expanding about $\delta=0$ gives
\begin{equation*}
h(L+L\delta)
=
h(L) 
+L h'(L) \delta 
+ \frac{L^2 h''(L)}{2} \delta^2
+ \frac{L^3 h'''(L)}{6} \delta^3
+ \frac{L^4 h''''(L)}{24} \delta^4
+ \O(\delta^5). 
\end{equation*}
Using the derivative attachment conditions and $\Lambda=L/H$, we obtain
\begin{equation}\label{poly-attachment-expansion}
\mathcal{P}(-2\delta-\delta^2)
=
1
-\frac{\Lambda^2}{2}\delta^2
-\frac{\Lambda^4}{8}\delta^4
+\mathcal{O}(\delta^5).
\end{equation}
Substituting $r=-2\delta-\delta^2$ into
\eqref{poly-general-form} and comparing with
\eqref{poly-attachment-expansion} yields
\begin{equation}\label{poly-fixed-coefficients}
c_1=0,
\qquad
c_2=-\frac{\Lambda^2}{8},
\qquad
c_3=-\frac{\Lambda^2}{16},
\qquad
c_4=-\frac{\Lambda^4+5\Lambda^2}{128}.
\end{equation}
The coefficients $c_5$ and $c_6$ remain undetermined. For convenience,
set
\begin{equation}\label{poly-free-coefficients}
A\coloneqq c_5+c_6,
\qquad
B\coloneqq-c_6.
\end{equation}
With this notation, \eqref{poly-general-form} takes the form
\eqref{ideal-polynomial-P}.

\subsection{Imposing the remaining constraints}

It remains to determine $A$ and $B$. The central-height condition
$\mathcal{P}(1)=\alpha$ in \eqref{poly-end-values} gives precisely the
coefficient $A$ in \eqref{ideal-polynomial-A}.
To impose the area constraint, we use the elementary identities
\begin{subequations}\label{poly-integrals}
\begin{gather}
\int_0^1(1-s^2)^2\dd{s}
=
\frac{8}{15},
\qquad
\int_0^1(1-s^2)^3\dd{s}
=
\frac{16}{35},
\qquad
\int_0^1(1-s^2)^4\dd{s}
=
\frac{128}{315},
\\
\int_0^1(1-s^2)^5\dd{s}
=
\frac{256}{693},
\qquad
\int_0^1s^2(1-s^2)^5\dd{s}
=
\frac{256}{9009}.
\end{gather}
\end{subequations}
Substituting \eqref{ideal-polynomial-P} into \eqref{poly-area} and
using \eqref{poly-integrals} gives
\begin{equation}\label{poly-area-equation}
\frac{\pi}{4\Lambda}
=
1
-\frac{\Lambda^2}{9}
-\frac{\Lambda^4}{315}
+\frac{256}{693}A
+\frac{256}{9009}B.
\end{equation}
Solving \eqref{poly-area-equation} for $B$ gives
\eqref{ideal-polynomial-B}. Thus the prescribed central height
determines $A$, after which the area constraint uniquely determines
$B$.

\subsection{Positivity and nondegeneracy}

Assume the parameter constraints given in
\eqref{ideal-polynomial-admissibility} and define
\begin{equation}\label{poly-D}
D
\coloneqq
\frac{\Lambda^4+19\Lambda^2}{32}
-5A+B.
\end{equation}
The second condition in
\eqref{ideal-polynomial-admissibility} gives $D>0$.
Differentiating \eqref{ideal-polynomial-h} gives
\begin{equation*}
h''(0)
=
-\frac{2H}{L^2}\mathcal{P}'(1).
\end{equation*}
From \eqref{ideal-polynomial-P}, we compute 
$\mathcal{P}'(1)=-D$, and therefore
\begin{equation}\label{poly-center-derivative}
h''(0)
=
\frac{2H}{L^2}D
>
0.
\end{equation}
Thus $D>0$ gives the required nondegeneracy at the center.

To establish positivity, write $\mathcal{P}$ in the degree-six
Bernstein basis:
\begin{equation}\label{poly-Bernstein}
\mathcal{P}(r)
=
\sum_{k=0}^6
\beta_k
\binom{6}{k}
r^k(1-r)^{6-k}.
\end{equation}
Conversion from the power-basis representation
\eqref{ideal-polynomial-P} gives
\begin{subequations}\label{poly-Bernstein-coefficients}
\begin{alignat}{3}
\beta_0
&=
\beta_1
=
1,
&\qquad
\beta_2
&=
1-\frac{\Lambda^2}{120},
&\qquad
\beta_3
&=
1-\frac{9\Lambda^2}{320},
\\
\beta_4
&=
1-\frac{25\Lambda^2}{384}
-\frac{\Lambda^4}{1920},
&
\beta_5
&=
\alpha+\frac{D}{6},
&
\beta_6
&=
\alpha.
\end{alignat}
\end{subequations}
The bound $1<\Lambda<5/2$ implies
$\beta_k>0$ for $0\leq k\leq4$, while $\alpha>0$ and $D>0$ imply
$\beta_5,\beta_6>0$. Since the Bernstein basis functions are
nonnegative on $[0,1]$ and sum to one,
\eqref{poly-Bernstein} gives $\mathcal{P}(r)>0$ for
$r\in[0,1]$. Consequently, $h(\xi_1)>0$ throughout $[-L,L]$.


\section*{Acknowledgments}

This work was supported by the Mark Kac Applied Mathematics
Postdoctoral Fellowship at the Center for Nonlinear Studies at
Los Alamos National Laboratory. OpenAI's ChatGPT was used to assist
with the algebraic derivation of the polynomial profile in
Section~\ref{app:polynomial-profile}. The calculation was independently
verified by direct substitution, and the author assumes full
responsibility for all content of the manuscript.

Los Alamos National Laboratory Report LA-UR-26-27597.


\bibliographystyle{siamplain}
\bibliography{references}

\end{document}